\documentstyle[11pt]{article}
\newcounter{subequation}[equation]
\input psfig.sty
\makeatletter

\def\thesubequation{\theequation\@alph\c@subequation}
\def\@subeqnnum{{\rm (\thesubequation)}}
\def\slabel#1{\@bsphack\if@filesw {\let\thepage\relax
   \xdef\@gtempa{\write\@auxout{\string
      \newlabel{#1}{{\thesubequation}{\thepage}}}}}\@gtempa
   \if@nobreak \ifvmode\nobreak\fi\fi\fi\@esphack}
\def\subeqnarray{\stepcounter{equation}
\let\@currentlabel=\theequation\global\c@subequation\@ne
\global\@eqnswtrue
\global\@eqcnt\z@\tabskip\@centering\let\\=\@subeqncr
$$\halign to \displaywidth\bgroup\@eqnsel\hskip\@centering
  $\displaystyle\tabskip\z@{##}$&\global\@eqcnt\@ne
  \hskip 2\arraycolsep \hfil${##}$\hfil
  &\global\@eqcnt\tw@ \hskip 2\arraycolsep
  $\displaystyle\tabskip\z@{##}$\hfil
   \tabskip\@centering&\llap{##}\tabskip\z@\cr}
\def\endsubeqnarray{\@@subeqncr\egroup
                     $$\global\@ignoretrue}
\def\@subeqncr{{\ifnum0=`}\fi\@ifstar{\global\@eqpen\@M
    \@ysubeqncr}{\global\@eqpen\interdisplaylinepenalty \@ysubeqncr}}
\def\@ysubeqncr{\@ifnextchar [{\@xsubeqncr}{\@xsubeqncr[\z@]}}
\def\@xsubeqncr[#1]{\ifnum0=`{\fi}\@@subeqncr
   \noalign{\penalty\@eqpen\vskip\jot\vskip #1\relax}}
\def\@@subeqncr{\let\@tempa\relax
    \ifcase\@eqcnt \def\@tempa{& & &}\or \def\@tempa{& &}
      \else \def\@tempa{&}\fi
     \@tempa \if@eqnsw\@subeqnnum\refstepcounter{subequation}\fi
     \global\@eqnswtrue\global\@eqcnt\z@\cr}
\let\@ssubeqncr=\@subeqncr
\@namedef{subeqnarray*}{\def\@subeqncr{\nonumber\@ssubeqncr}\subeqnarray}
\@namedef{endsubeqnarray*}{\global\advance\c@equation\m@ne%
                           \nonumber\endsubeqnarray}

\makeatletter
\@addtoreset{equation}{section}
\makeatother
\renewcommand{\theequation}{\thesection.\arabic{equation}}

\def\dalemb#1#2{{\vbox{\hrule height .#2pt
        \hbox{\vrule width.#2pt height#1pt \kern#1pt
                \vrule width.#2pt}
        \hrule height.#2pt}}}
\def\square{\mathord{\dalemb{6.8}{7}\hbox{\hskip1pt}}}

\let\a=\alpha \let\b=\beta   \let\e=\epsilon
  \let\q=\theta  
  \let\n=\nu

\def\nn{\nonumber} \def\bd{\begin{document}} \def\ed{\end{document}}
\def\ds{\documentstyle} \let\fr=\frac \let\bl=\bigl \let\br=\bigr
\let\Br=\Bigr \let\Bl=\Bigl
\let\bm=\bibitem
\let\na=\nabla
\let\pa=\partial \let\ov=\overline
\def\ie{{\it i.e.\ }}
\def\etc{{\it etc.\ }}
\newcommand{\be}{\begin{equation}}
\newcommand{\ee}{\end{equation}}
\def\ba{\begin{array}}
\def\ea{\end{array}}
\def\ft#1#2{{\textstyle{{\scriptstyle #1}\over {\scriptstyle #2}}}}
\def\fft#1#2{{#1 \over #2}}
\def\del{\partial}
\def\sst#1{{\scriptscriptstyle #1}}
\def\oneone{\rlap 1\mkern4mu{\rm l}}
\def\e7{E_{7(+7)}}
\def\td{\tilde}
\def\wtd{\widetilde}
\def\im{{\rm i}}
\def\bog{Bogomol'nyi\ }
\def\q{{\tilde q}}
\def\hast{{\hat\ast}}
\def\0{{\sst{(0)}}}
\def\1{{\sst{(1)}}}
\def\2{{\sst{(2)}}}
\def\3{{\sst{(3)}}}
\def\4{{\sst{(4)}}}
\def\5{{\sst{(5)}}}
\def\6{{\sst{(6)}}}
\def\7{{\sst{(7)}}}
\def\8{{\sst{(8)}}}
\def\n{{\sst{(n)}}}
\def\oo{{\"o}}
\def\hA{\hat{\cal A}}
\def\ns{{\sst {\rm NS}}}
\def\rr{{\sst {\rm RR}}}
\def\tH{{\widetilde H}}
\def\tB{{\widetilde B}}
\def\cA{{\cal A}}
\def\cF{{\cal F}}
\def\tF{{\wtd F}}
\def\Z{\rlap{\sf Z}\mkern3mu{\sf Z}}
\def\ep{{\epsilon}}
\def\IIA{{\rm IIA}}
\def\IIB{{\rm IIB}}
\def\ads{{\rm AdS}}
\def\R{\rlap{\rm I}\mkern3mu{\rm R}}

\def\Ei{{\hbox{Ei}}}
\def\Ci{{\hbox{Ci}}}
\def\Si{{\hbox{Si}}}

\def\comment#1{  \begin{raggedright}{\tt [#1]}\end{raggedright}}

\newcommand{\ho}[1]{$\, ^{#1}$}
\newcommand{\hoch}[1]{$\, ^{#1}$}
\newcommand{\bea}{\begin{eqnarray}}
\newcommand{\eea}{\end{eqnarray}}
\newcommand{\ra}{\rightarrow}
\newcommand{\lra}{\longrightarrow}
\newcommand{\Lra}{\Leftrightarrow}
\newcommand{\ap}{\alpha^\prime}
\newcommand{\bp}{\tilde \beta^\prime}
\newcommand{\tr}{{\rm tr} }
\newcommand{\Tr}{{\rm Tr} }
\newcommand{\NP}{Nucl. Phys. }
\newcommand{\tamphys}{\it Center for Theoretical Physics,
Texas A\&M University, College Station, TX 77843}
\newcommand{\upenn}{\it Dept. of Phys. and Astro.,
University of Pennsylvania,
Philadelphia, PA 19104}

\newcommand{\auth}{M. Cveti\v{c}\hoch{\dagger\$1},
S.S. Gubser\hoch{\star2}, H. L\"u\hoch{\dagger1} and
C.N. Pope\hoch{\ddagger *3} }

\begin{document}
\begin{flushright}
\hfill{
UPR/0856-T \ \ \ \ 
CTP TAMU-38/99 \ \ \ \ 
HUTP-99/A049 \ \ \ \ 
NSF-ITP-99-106}\\
\hfill{\bf hep-th/9909121}\\
\hfill{September 1999}\\

\end{flushright}


\begin{center}

{\large {\bf Symmetric Potentials of Gauged Supergravities in 
Diverse Dimensions\\
and Coulomb Branch of Gauge Theories}}

\vspace{20pt}

\auth

\vspace{10pt}
{\hoch{\dagger}\upenn}

\vspace{10pt}
{\hoch{\ddagger}\tamphys}

\vspace{10pt}
{\hoch{\star}{\it Lyman Laboratory of Physics, Harvard University,
Cambridge, MA 02138} }

\vspace{10pt}
{\hoch{\$}{\it Institute for Theor. Phys., Univ. of Santa
Barbara, Santa Barbara, CA 93106}}

\vspace{10pt}
{\hoch{*}{\it SISSA, Via Beirut, No. 2-4, 34013 Trieste,
Italy}}

\vspace{30pt}

\underline{ABSTRACT}
\end{center}

      A class of conformally flat and asymptotically anti-de Sitter
geometries involving profiles of scalar fields is studied from the
point of view of gauged supergravity.  The scalars involved in the
solutions parameterise the $SL(N,\R)/SO(N)$ submanifold of the full
scalar coset of the gauged supergravity, and are described by a
symmetric potential with a universal form.  These geometries descend
via consistent truncation from distributions of D3-branes, M2-branes,
or M5-branes in ten or eleven dimensions.  We exhibit analogous
solutions asymptotic to AdS$_6$ which descend from the D4-D8-brane
system. We obtain the related six-dimensional theory by consistent
reduction from massive type IIA supergravity.  All our geometries
correspond to states in the Coulomb branch of the dual conformal field
theories.  We analyze linear fluctuations of minimally coupled scalars
and find both discrete and continuous spectra, but always bounded
below.


\pagebreak
\setcounter{page}{1}

\tableofcontents
\addtocontents{toc}{\protect\setcounter{tocdepth}{2}}
\newpage

\section{Introduction}

The AdS/CFT correspondence \cite{juanAdS,gkPol,witHolOne} offers the
possibility of extracting information about gauge theory in a strongly
coupled limit from a dual supergravity description.  Much recent work (see
for instance \cite{GPPZ1,DZ,fgpw1,fgpw2,behrndt}) has been devoted to bulk
geometries that are only asymptotically anti-de Sitter and retain
Poincar\'e invariance in the boundary directions.  Such geometries
correspond either to a relevant deformation of the conformal field theory
or a state of the conformal field theory with a VEV which sets an energy
scale.  The geometries arise from solving a system of scalars coupled to
gravity.  At the linearized level, the second order scalar equations admit
two linearly independent solutions.  Modulo the subtleties discussed in
\cite{klebWit}, a bulk geometry corresponds to a deformation of the
conformal field theory if the scalar profiles approach the more singular of
the two possible linearized solutions near the boundary, and it corresponds
to a state of the conformal field theory if they approach the less singular
solutions.

We shall be concerned in this paper with fairly broad class of
asymptotically anti-de Sitter solutions which can nevertheless be
obtained in closed form.  These geometries lift to solutions of ten or
eleven-dimensional supergravity describing some distribution of
parallel D3-branes, M2-branes, or M5-branes.  They all preserve one
half of supersymmetry, and they are states of the dual conformal field
theories: the VEV's in question are simply the positions of the branes
in the distribution.

These geometries are interesting for several reasons.  First, a subset
of them arise as limits of charged black holes in anti-de Sitter
space.  The limit in question leads (generically) to a naked timelike
singularity.  When the asymptotically anti-de Sitter solution is
lifted to ten or eleven dimensions, the only singularities are due to
the fact that the branes are distributed continuously.  They are naked
null singularities similar to those around low-dimensional D-branes.

    Second, the geometries provide examples of consistent truncation
in various dimensions.  The explicit embedding of lower dimensional
gauged supergravities in higher dimensions has been worked out only
for a handful of cases, namely $N=2$ (maximal) and $N=1$ gauged $D=7$
supergravity as $S^4$ reductions from $D=11$ \cite{NieuNew,d7gauge},
and $D=6$ gauged $N=2$ supergravity as as a warped $S^4$ reduction
from massive type IIA supergravity \cite{d6gauge}. (For the $S^7$
reduction of $D=11$ supergravity, the results of \cite{deWit1,deWit2}
are complete in principle but very implicit in form.)  Consistent
embeddings in $D=11$ and $D=10$ type IIB for the gauged supergravities
with the maximal abelian subgroups $U(1)^4$, $U(1)^2$ and $U(1)^3$ of
the $SO(8)$, $SO(5)$ and $SO(6)$ gauge groups from the maximal $S^7$,
$S^4$ and $S^5$ reductions have also been constructed, in \cite{ten}.
It is therefore of interest in the ongoing study of consistent
truncations to have further examples of some generality.

Some of the geometries we consider appeared in \cite{KLT,rosf}, and a set
of representative examples was studied in \cite{fgpw2}.  In \cite{fgpw2}
the curious result was obtained that some of the asymptotically anti-de
Sitter geometries lifted to distributions of branes of indefinite sign:
that is, there were ``branes'' of negative charge and negative tension
present in the distribution.  Negative charge and positive tension would
indicate an anti-brane.  But negative charge and negative tension, in the
absence of some orientifolding, is pathological.  For instance, the
transverse fluctuations of such branes would have negative kinetic terms.
Yet the asymptotically anti-de Sitter geometries that lead to these
disallowed brane distributions are no more pathological than those where
the brane distribution is strictly positive.

The paper is organized as follows.  In
section~\ref{SymmetricPotentials}, we describe the interacting scalars
which participate in the conformally flat and asymptotically anti-de
Sitter geometries.  Because of the similarities of these geometries
with supergravity domain wall solutions (see for example
\cite{CveticSoleng}), we will often refer to them as anti-de Sitter
domain walls.  In section~\ref{HigherDimensional} we explain the ten
and eleven dimensional origin of the geometries.  In
section~\ref{pBraneOrigin} we describe explicitly the brane
distributions corresponding to each type of domain wall.  In
section~\ref{AdS6}, we consider asymptotically $\hbox{AdS}_6$
geometries which arise from many D4-branes distributed inside
coincident D8-branes.  We will see that D3-branes, M2-branes, and
M5-branes behave quite similarly in their respective dimensions, and
the geometries and distributions can be described very much in
parallel.  The D4-D8-brane solutions are rather different and provide
another interesting example of consistent truncation.  Finally, in
section~\ref{Spectrum}, we analyze the spectrum of minimally coupled
scalars in the various domain wall geometries we have constructed.

\section{Symmetric Potentials in Gauged Supergravities, \\
and Domain-wall Solutions}
\label{SymmetricPotentials}

   In this section, we consider the structure of the scalar potentials
that arise in the maximally-supersymmetric gauged supergravities in
$D=4$, 5 and 7 dimensions.  We shall present the results in a slightly
more general framework for arbitrary dimension $D$, although in the
end we shall see that $D=4$, 5 and 7 are singled out.

\subsection{Symmetric potentials}

    The scalar fields in $D$-dimensional maximal supergravity
parameterize the coset $E_{11-{\sst D}}/K$, where $E_n$ is the
maximally-non-compact form of the exceptional group $E_n$, and $K$ is
its maximal compact subgroup.\footnote{As usual, we define $E_5\sim
D_5$, $E_4\sim A_4$, $E_3\sim A_2\times A_1$, $E_2\sim A_1\times \R$
and $E_1\sim \R$.}  We can focus on the $SL(N,\R)$ subgroup of $E_n$,
and consider in particular the subset of $\ft12 N(N+1)-1$ scalars
parameterizing the coset $SL(N,\R)/SO(N)$.  (In $D=4$ and $D=5$, the
relevant subgroups of $E_7$ and $E_6$ are $SL(8,\R)$ and $SL(6,\R)$.
In $D=7$, the group $E_4$ is itself just $A_4$, and so we consider
$SL(5,\R)$ in that case.)

   One can use the local $SO(N)$ transformations in order to
diagonalize the scalar potential.  Thus we are led to consider the
following Lagrangian for gravity plus scalars in $D$ dimensions:
\be
e^{-1}\, {\cal L}_D = R -\ft12 (\del\vec\varphi)^2 - V\,,\label{ddlag}
\ee
where the potential $V$ is given by
\be
V = -\ft12 g^2\, \Big( (\sum_{i=1}^N
 X_i)^2 - 2 \sum_{i=1}^N X_i^2 \Big)\,.\label{ddpot}
\ee
(In $D=4$, 5 and 7, we shall have $N=8$, 6 and 5 respectively.)  The
$N$ quantities $X_i$, which are subject to the constraint
\be
\prod_{i=1}^N X_i = 1\,,\label{prodcon}
\ee
can be parameterized in terms of $(N-1)$ independent
dilatonic scalars $\vec\varphi$ as follows:
\be
X_i = e^{-\fft12 \vec b_i\cdot\vec\varphi}\,,\label{ddxdef}
\ee
where the $\vec b_i$ are the weight vectors of the fundamental
representation of $SL(N,\R)$, satisfying
\be
\vec b_i\cdot\vec b_j = 8\delta_{ij} - \fft{8}{N}\,,\qquad
\sum_i \vec b_i=0\,,\qquad
(\vec u\cdot\vec b_i)\, \vec b_i = 8 \vec u\,,\label{dotprod}
\ee
where $\vec u$ is an arbitrary $N$-vector.
The last equation in (\ref{dotprod}) allows us to express the dilatons
$\vec\varphi$ in terms of the $X_i$:
\be
\vec\varphi = -\ft14 \sum_i \vec b_i\, \log X_i\,.\label{phiexp}
\ee
Note that the potential has a minimum at $X_i=1$ for $N\le 3$, a point
of inflection at $X_i=1$ for $N=4$, and a maximum at $X_i=1$ for
$N\ge 5$.

       The equations of motion for the scalar fields, following from
(\ref{ddlag}), are
\be
\square \vec\varphi = \fft{\del V}{\del\vec\varphi}\,.\label{ddeom}
\ee
 From (\ref{ddxdef}) it follows that $\del X_i/\del\vec\varphi = -\ft12
\vec b_i\, X_i$, and hence the equations of motion (\ref{ddeom})
become
\be
\square\vec\varphi = \ft12 g^2\, \sum_i \vec b_i
\Big( X_i\, \sum_j X_j  - 2  X_i^2\Big)\,.\label{phieq}
\ee
Note that we can also write the scalar equations of motion as
\be
\square\log X_i = 2 g^2\, \Big( 2 X_i^2 - X_i\, \sum_j X_j - \ft2{N}
\sum_j X_j^2 + \ft1{N} (\sum_j X_j)^2 \Big)\,.\label{xeq0}
\ee
(The last two terms on the right-hand side imply that the sum over $i$
is zero, as it must be since the $N$ quantities $X_i$ satisfy
(\ref{prodcon}).)

   In order to make contact with previous results, we can verify that
if the potential (\ref{ddpot}) is specialized to the case where as
many pairs of the quantities $X_i$ as possible are set equal, we
recover the previously-known potentials that arise when the various
gauged supergravities are truncated to the multiplets comprising the
maximal abelian subgroups of the original gauge groups. Thus for $D=4$
we may set the 8 quantities $X_i$ pairwise equal,
\be
X_1=X_2\equiv \wtd X_1\,,\quad 
X_3=X_4\equiv \wtd X_2\,,\quad
X_5=X_6\equiv \wtd X_3\,,\quad
X_7=X_8\equiv \wtd X_4\,,
\ee
whereupon the potential becomes
\be
V =  -4 g^2\, \sum_{a<b} \wtd X_a\, \wtd X_b\,,\label{d4oldpot}
\ee
where the four $\wtd X_a$ satisfy $\prod_a \wtd X_a=1$.  This
potential is the one encountered in \cite{duffliu,ten}, for the
scalars in the truncation of four-dimensional gauged $SO(8)$
supergravity to its $U(1)^4$ maximal abelian subgroup.

    For $D=5$, we may set the 6 quantities $X_i$ pairwise equal, 
\be
X_1=X_2\equiv \wtd X_1\,, \qquad X_3=X_4\equiv \wtd X_2\,,\qquad
X_5=X_6 \equiv \wtd X_3\,,\label{d5xtoy}
\ee
under which the potential $V$ given in (\ref{ddpot}) reduces to
\be
V= -4g^2\, \sum_{a=1}^3  \wtd X_a^{-1}\,,\label{d5oldpot}
\ee
where the three scalar quantities $\wtd X_a$ satisfy $\prod \wtd
X_a=1$.  This potential was encountered in \cite{kcs,ten}; it arises
in the truncation of the five-dimensional gauged $SO(6)$ supergravity
to its $U(1)^3$ maximal abelian subgroup.

   Finally, in $D=7$ we may set 
\be
X_1=X_2\equiv \wtd X_1\,,\quad
X_3=X_4\equiv \wtd X_2\,,\quad
X_5 \equiv \wtd X_0\,.
\ee
The three $\wtd X_a$ satisfy $\wtd X_0 = (\wtd X_1\, \wtd X_2)^{-2}$,
and the potential (\ref{ddpot}) becomes
\be
V = -\ft12 g^2( -\wtd X_0^2 + 8 \wtd X_1\, \wtd X_2 + 4  \wtd X_0\, \wtd X_1
 +  4  \wtd X_0\, \wtd X_2)\,.\label{d7oldpot}
\ee
This can be recognized as precisely the potential previously
encountered in the truncation of seven-dimensional $SO(5)$ gauged
supergravity to its $U(1)^2$ maximal abelian subgroup \cite{ten}.  It is
interesting to observe how the ``asymmetry'' between the $\wtd X_0$
field and the $\wtd X_1$ and $\wtd X_2$ fields here can be simply
understood as originating from the fact that the more general
potential (\ref{ddpot}) has an odd number of fields $X_i$ in this
case, and so there is inevitably an ``odd one out'' after setting as
many pairs as possible equal.

    It should be emphasized that in each case, setting pairs of $X_i$
equal is consistent with the equations of motion (\ref{xeq0}).

\subsection{AdS domain-wall solutions}

    The equations of motion following from the gravity plus scalar
Lagrangian (\ref{ddlag}) comprise the scalar equations (\ref{phieq}),
together with the Einstein equation
\be
R_{MN} = \ft12 \del_M\vec\varphi\cdot \del_N\vec\varphi + \ft1{D-2}\,
V\,  g_{MN}\,.\label{einsteq}
\ee
We find that these equations admit solutions given by
\bea
ds_D^2 &=& (g\, r)^{\fft{4}{D-3}}
\, (\prod_i H_i)^{\fft12 - \fft{2}{N}}\, dx^\mu\, dx_\mu +
 (\prod_i H_i)^{-\fft2{N}}\, \fft{dr^2}{g^2\, r^2}\,,\nn\\
X_i &=& H_i^{-1}\,  (\prod_j H_j)^{\fft1{N}}\,,\label{ddsol}
\eea
where
\be
H_i = 1 +\fft{\ell_i^2}{r^2}\,,\label{hform}
\ee
provided that the integer $N$ is related to the spacetime dimension
$D$ by
\be
N = \fft{4(D-2)}{D-3}\,.\label{ndrel}
\ee
Note that for $D=4$, 5 and 7 this indeed implies $N=8$, 6 and 5
respectively.  As we explained in the Introduction, we shall refer to
solutions of this kind as anti-de Sitter domain walls.

    In verifying these solutions, it is useful first to calculate
the Ricci tensor for the class of metrics
\be
ds_D^2 = e^{2A}\, dx^\mu\, dx_\mu + e^{2B}\, dr^2\,,
\ee
where $A$ and $B$ are functions of $r$.  In the orthonormal basis
$e^\mu= e^A\, dx^\mu$, $e^r=e^B\, dr$, we find that the tangent-space
components of the Ricci tensor are
\bea
R_{\mu\nu} &=& - (A'' + (D-1)\, {A'}^2 - A'\, B')\, e^{-2B}\,
\eta_{\mu\nu}\,,\nn\\
R_{rr} &=& - (D-1)\, (A'' + {A'}^2 - A'\, B')\, e^{-2B}\,,
\eea
where a prime denotes a derivative with respect to $r$.
Thus the Einstein equation (\ref{einsteq}) becomes
\bea
 -(A'' + (D-1)\, {A'}^2 - A'\, B')\, e^{-2B}&=& \ft1{D-2} V\,,\nn\\
-(D-1)(A'' + {A'}^2 - A'\, B')\, e^{-2B}&=& 
\ft12 e^{-2B}\, \vec\varphi'\cdot \vec\varphi' +
 \ft1{D-2}\,  V\,.\label{ddeinst}
\eea
It is useful also to note that (\ref{hform}) implies $H_i' = (-2/r)\,
(H_i-1)$.  Substituting into the Einstein and scalar equations of
motion, we now straightforwardly verify that (\ref{ddsol}) is a
solution.

   It should be noted that the solutions we have obtained here reduce
to previously-known ones in certain special cases.  Namely, if we make
the pairwise identifications $\ell_{2p-1}=\ell_{2p}\equiv L_p$ for
$1\le p\le N/2$ in $D=4$ or $D=5$, then we obtain solutions with 4 or
3 parameters $L_p$ respectively, which correspond precisely to the
extremal BPS limits of the 4-charge \cite{duffliu,ten} and 3-charge
AdS$_4$ \cite{kcs} and AdS$_5$ black holes of maximal $D=4$ and $D=5$
supergravity.  In the $D=7$ case, if we set $\ell_1=\ell_2\equiv L_1$,
$\ell_3=\ell_4\equiv L_2$, and $\ell_5=0$, we recover the
previously-known 2-parameter solutions that arise as the extremal BPS
limits of the 2-charge AdS$_7$ black holes \cite{ten} of maximal
gauged $D=7$ supergravity.\footnote{In all cases the charges vanish
when the extremal limit is taken.} 

   It should be emphasized also that although we naively appear to
have $N$ parameters $\ell_i$ in the general solutions (\ref{ddsol}),
there are really only $(N-1)$ genuinely independent ones.  To see
this, let us suppose, without loss of generality, that $\ell_N^2$ is
the smallest of all the parameters, \ie $\ell_i^2\ge \ell_N^2$.  We
also decompose the index $i$ into $i=(a,N)$, where $1\le a\le N-1$.
We now make the following coordinate transformation,
\be
r^2 = R^2 - \ell_N^2\,,\label{rrtrans}
\ee
and at the same time define
\be
\wtd H_a = 1 + \fft{\lambda_a^2}{R^2}\,, \qquad
\lambda_a^2 \equiv \ell_a^2 - \ell_N^2\,.
\ee
Straightforward algebra now shows that in terms of these quantities,
the metric in the solution (\ref{ddsol}) becomes
\be
ds_D^2 = (g\, R)^{\fft{4}{D-3}}
\, (\prod_a \wtd H_a)^{\fft12 - \fft{2}{N}}\, dx^\mu\, dx_\mu +
 (\prod_a \wtd H_a)^{-\fft2{N}}\, \fft{dR^2}{g^2\, R^2}\,,
\ee
where the products are now only over the $(N-1)$ functions $\wtd H_a$.
The quantities $X_i$ in (\ref{ddsol}) become
\be
X_a= \wtd H_a^{-1}\, (\prod_b \wtd H_b)^{\fft1{N}}\,,\qquad
X_N =  (\prod_b \wtd H_b)^{\fft1{N}}\,.
\ee
Thus we see that after making the coordinate transformation
(\ref{rrtrans}), the solution (\ref{ddsol}) with $N$ parameters
$\ell_i^2$ is nothing but a solution of the identical form, but with
transformed parameters
\be
\ell_i^2=(\ell_1^2,\ell_2^2,\ldots,\ell_{N-1}^2,  \ell_N^2) \longrightarrow 
(\ell_1^2-\ell_N^2, \ell_2^2-\ell_N^2,\ldots, \ell_{N-1}^2-\ell_N^2,
0)\,.\label{ellTransform}
\ee
Thus there are actually only $(N-1)$ independent parameters in the
solutions (\ref{ddsol}), implying 7, 6 and 4 in $D=4$, 5 and 7
respectively.  

A simple example of the transformation (\ref{ellTransform}) is the
case where all the $\ell_i$ have a common value, $\ell$.  In this case
the above transformation leads to a solution where all the $\ell_i$ vanish,
and the the geometry is locally exactly anti-de Sitter.  It is
possible however for the geometry to change abruptly at some radius.
This is the case, for example, if one has a spherical shell of
D3-branes at a radius $\ell$.  Outside the shell, the geometry is
$\hbox{AdS}_5 \times S^5$, exactly the same as if the shell were
collapsed to a point.  Inside the shell the geometry is
ten-dimensional flat space.  This configuration has been considered in
\cite{KLT,ChepelevRoiban,GiddingsRoss}.  It was pointed out
already in \cite{KLT} and emphasized in \cite{GiddingsRoss} that
the background outside the shell is insensitive to what spherical
distribution of branes one puts inside the shell: only the total
number of branes controls the overall curvature.

The transformation (\ref{ellTransform}) is a generalized
shell theorem: as we will see in later sections, it effectively
exhibits a class of brane distributions that all have the same
exterior.

    Further applications of the transformation (\ref{ellTransform})
arise if we consider a situation where $m$ of the $\ell_i^2$ parameters
are equal.  Suppose we order the parameters so that $\ell_{N-m+1}^2=
\ell_{N-m+2}^2 =\cdots = \ell_N^2\equiv L^2$.  Then, after making the
transformation (\ref{ellTransform}), we will have new parameters
\be
\td\ell_i^2 = (\td\ell_1^2, \td \ell_2^2,\ldots,
\td\ell_{N-m}^2,0,0,\ldots 0)\,,
\ee
where $\td\ell_i^2 = \ell_i^2-L^2$, and so the configuration is mapped
to one with only $(N-m)$ non-vanishing parameters.  If $L^2$ is the
smallest value among the original parameters, then all the transformed
parameters $\td\ell_i^2$ will be positive.  However, if some of the original
parameters were smaller than $L^2$, then after the transformation they
will correspond to $\td\ell_i^2$ parameters that are negative.

    In fact, quite generally we can use the transformation to map
the sets of solutions with purely non-negative $\ell_i^2$ parameters into
sets of solutions with arbitrary numbers $p$ and $q$ of positive and
negative $\ell_i^2$, with $0\le p+q\le N$.  For example, the solution
with $\ell_2^2=\ell_3^2=\cdots =\ell_N^2\equiv L^2>0$ and $\ell_1^2=0$
transforms into a solution with $\td\ell_i^2=(-L^2,0,0,\ldots,0)$,
with $\td\ell_1^2<0$.  Thus a special case of the $n=N-1$ solutions
with positive $\ell_i^2$ transforms to an $n=1$ solution with negative
$\ell_1^2$.  Likewise, we can obtain a solution with $n$ negative
$\ell_i^2$ parameters from a solution with $(N-n)$ positive parameters.
These transformations will be useful later, when we
derive results for the charge distributions that allow us to view the
domain-wall solutions oxidized to higher dimensions as distributions
of M-branes or D-branes.   Thus having obtained results for the
distributions for arbitrary numbers $n$ of positive $\ell_i^2$
parameters, we can immediately read off results for sets of negative
parameters.  This mapping between solutions with positive $\ell_i^2$
parameters and negative parameters will also be useful when we
consider the spectra of wave equations in these background geometries
in section 6.  In general, for reasons explained in section 6.1, we
refer to parameters with $\ell_i^2>0$ as ``Lorentzian,'' and
parameters with $\ell_i^2<0$ as ``Euclidean.''

\section{Higher-dimensional Origin}
\label{HigherDimensional}

    In the previous section, we considered the $D$-dimensional
theories of gravity plus scalar fields that arise as consistent
truncations of the gauged maximal supergravities in $D=4$, 5 and 7,
and we showed that they admit multi-parameter domain-wall solutions.
It is believed that the gauged maximal supergravities can all be
obtained as consistent Kaluza-Klein reductions using sphere
compactifications of certain higher-dimensional supergravities.
Specifically, it has been shown that the $D=4$ and $D=7$ gauged
theories arise from the $S^7$ and $S^4$ reductions of
eleven-dimensional supergravity, while the $D=5$ gauged theory is
expected to arise from the $S^5$ reduction of type IIB
supergravity.\footnote{It has also been shown that the largest gauged
supergravity in $D=6$ arises as an $S^4$ reduction of massive type IIA
supergravity \cite{d6gauge}.}  We can therefore expect that the
truncations of the various gauged supergravities that we considered in
the previous section should themselves be directly derivable as
consistent Kaluza-Klein reductions from the relevant
higher-dimensional supergravities.

   In this section, we shall show explicitly how the various
$D$-dimensional theories that we discussed in section 2 can be
embedded into the higher-dimensional supergravities.  The crucial
point will be that these are {\it consistent} embeddings, meaning that
all solutions of the lower-dimensional theories will be solutions of
the higher-dimensional ones.  In each case, the scalar fields of the
lower-dimensional theory will appear as parameters describing
inhomogeneous deformations of the compactifying sphere metric.
Because of the inhomogeneity, demonstrating the {\it consistency} of
the reduction procedure is highly non-trivial, since there is no
simple group-invariance argument that can account for it.  In fact a
crucial point in all such Kaluza-Klein reductions is that the scalar
fields appear not only in the metric reduction Ansatz, but also in the
reduction Ansatz for an antisymmetric tensor field of the
higher-dimensional supergravity.  It is only because of remarkable
``conspiracies'' between the contributions from the scalars in the
metric and field strength Ans\"atze that the consistent reduction is
possible.  Thus the issue of consistency is one that can be addressed
only if the complete reduction Ans\"atze, including those for the
antisymmetric tensor fields, are given.

   Let us first present our results for the consistent reduction
Ans\"atze.  We can give all three cases, for the reductions to $D=4$,
5 and 7, in a single set of formulae.  In each case the relevant part
of the higher-dimensional supergravity theory comprises the metric
$d\hat s^2$, and an antisymmetric tensor field strength $\hat
F_{\sst{(D)}}$.\footnote{We place hats on all higher-dimensional
fields, and also on the Hodge $*$ operator in the higher dimension.}
For the case of $D=4$, we have the 4-form field $\hat F_\4$ of
eleven-dimensional supergravity.  For $D=7$, we have $\hat F_\7={\hat
*\hat F_\4}$, where $\hat F_\4$ is again the 4-form of
eleven-dimensional supergravity.  For $D=5$, we have $\hat F_\5$, from
which the self-dual 5-form $\hat H_\5$ will be constructed as $\hat
H_\5 = \hat F_\5 + {\hat *\hat F_\5}$.  Our expressions for the metric
and field-strength Ans\"atze are
\bea
d\hat s^2 &=& \wtd\Delta^{\fft2{D-1}} \, ds_D^2 + \fft1{g^2}\,
\wtd\Delta^{-\fft{D-3}{D-1}}\, \sum_i X_i^{-1}\, d\mu_i^2\,,\nn\\
\hat F_{\sst{(D)}} &=& g\, \sum_i(2 X_i^2\, \mu_i^2 - \wtd\Delta\, X_i)\,
\ep_{\sst{(D)}}
-\fft1{2g}\, \sum_i X_i^{-1} \, {* dX_i}\wedge d(\mu_i^2)\,,\label{dgenansatz}
\eea
where
\be
\wtd\Delta = \sum_i X_i\, \mu_i^2\,,
\ee
and the $\mu_i$ are a set of $N$ ``direction cosines'' that satisfy
the constraint
\be
\sum_i \mu_i^2 = 1\,.\label{mucon}
\ee
In (\ref{dgenansatz}), $\ep_{\sst{(D)}}$ denotes the volume form of the
$D$-dimensional metric $ds_D^2$.  

    Clearly, if the scalar fields $\vec\varphi$ all vanish, so that we
have $X_i=1$, the internal metric in (\ref{dgenansatz}) reduces to the
standard round metric on the $(N-1)$-sphere, proportional to $\sum_i
d\mu_i^2$.  When the scalars are nonzero, they parameterize
inhomogeneous deformations of the round sphere metric.  The scalars
also appear in the Ansatz for the antisymmetric tensor, as given in
the second line in (\ref{dgenansatz}).  It is straightforward to show
that with this Ansatz, the equations of motion and Bianchi identity
for the field strength $\hat F_{\sst{(D)}}$ correctly reproduce the
$D$-dimensional scalar equations of motion that we derived in section 2.  In
fact it is not hard to see that in each of the cases $D=4$, 5 and 7,
the equations of motion for the scalars come from the Bianchi identity
\be
d\hat F_{\sst{(D)}}=0\,.\label{bianchi}
\ee
For the case $D=4$, (\ref{bianchi}) is, of course, directly the
Bianchi identity of the 4-form field of $D=11$ supergravity.  For
$D=7$, where we have $\hat F_\7={\hat* \hat F_\4}$, the Bianchi
identity (\ref{bianchi}) originates from the equation of motion for
$\hat F_\4$, namely $d{\hat *\hat F_\4} = \ft12 \hat F_\4\wedge \hat
F_\4$.  Since the Ansatz for $\hat F_\7$ in (\ref{dgenansatz}) implies
that $ \hat F_\4\wedge \hat F_\4=0$, we are again just left with
(\ref{bianchi}).   Finally, in $D=5$ the Bianchi identity for the
self-dual 5-form $\hat H_\5 = \hat F_\5 + {\hat *\hat F_\5}$ is just 
$d\hat H_\5=0$, since we are taking the other fields of the type IIB
theory to be zero in our reduction.  Furthermore, from the form of the
Ansatz for $\hat F_\5$ given in (\ref{dgenansatz}) in this case, we
can see that we will separately have to have $d\hat F_\5=0$ and
$d{\hat * \hat F_\5}=0$, and in fact it is the former that will imply
the equations of motion for the $X_i$.  In fact in all three cases,
the ``field equation'' $d{\hat * \hat F_{\sst{(D)}}}=0$ can be seen to
be identically satisfied, supplying no further information.

    To see how (\ref{bianchi}) implies the equations of motion for the
lower-dimensional scalars, we note from the Ansatz for $\hat
F_{\sst{(D)}}$ in (\ref{dgenansatz}) that the
only terms arising will be when the exterior derivative lands on the
$\mu_i$ coordinates in the prefactor of $\ep_{\sst{(D)}}$, and on the
$X_i^{-1}\, {*dX_i}$ factor in the last term in the field-strength
Ansatz.  Thus we deduce that
\be
\sum_i\Big( \square \log X_i - 4 g^2\, X_i^2 + 2 g^2\, X_i\, \sum_j
X_j \Big)\, d(\mu_i^2) =0\,.\label{xeq1}
\ee
We cannot simply deduce from this that the factor enclosed in the
large parentheses vanishes for each $i$, since the $\mu_i$ coordinates
are not independent, but are subject to the constraint (\ref{mucon}),
implying that $\sum_i d(\mu_i^2) =0$.  Thus the most we can deduce from  
(\ref{xeq1}) is that
\be
\square\log X_i =  4 g^2\, X_i^2 - 2 g^2\, X_i\, \sum_j X_j + Q\,,
\label{xeq2}
\ee
where $Q$ is an undetermined $i$-independent quantity.  However, since
we know that the $X_i$ satisfy (\ref{prodcon}), it follows that
$\sum_i \log X_i=0$.  Thus summing over $i$ in (\ref{xeq2}), the
left-hand side must vanish, which gives us an equation for $Q$.
Substituting back into (\ref{xeq2}) then gives
\be
\square\log X_i = 2 g^2\, \Big( 2 X_i^2 - X_i\, \sum_j X_j - \ft2{N}
\sum_j X_j^2 + \ft1{N} (\sum_j X_j)^2 \Big)\,,
\ee
which is precisely the scalar equations of motion that we derived for
the Lagrangians in section 2.

    A complete verification of the consistency of the reduction Ansatz
(\ref{dgenansatz}) would require substituting it into the
higher-dimensional Einstein equation.  The components of this equation
with indices living in the lower dimensions should then give rise to
the $D$-dimensional Einstein equation (\ref{ddeinst}), while the
internal and mixed components should give rise to the scalar equations
of motion, and to certain non-trivial self-consistency checks.  We
have not yet performed a complete check of the Einstein equation, on
account of the complexity of the curvature calculations that are
involved here.  However, the computations here are rather similar to
ones that arose in \cite{ten}, where a number of non-trivial
checks on the consistency of the Ansatz in the Einstein equation were
performed.  In fact, if we specialize our Ans\"atze here by setting
as many pairs of $X_i$'s as possible equal, they can be seen to reduce
precisely to special cases of the Ans\"atze considered in \cite{ten}.

   To see how this occurs, consider first a case where $N$ is even.
Then, we may set 
\be
X_1=X_2\equiv \wtd X_1\,,\qquad  X_3 =X_4\equiv \wtd X_2\,,
\qquad \hbox{\etc}\label{xpairs}
\ee
and also introduce angles $\phi_a$ such that
\bea
&&\mu_1=\td\mu_1\, \cos\phi_1\,,\qquad
\mu_2=\td\mu_1\, \sin\phi_1\,,\nn\\
&&\mu_3=\td\mu_2\, \cos\phi_2\,,\qquad
\mu_4=\td\mu_2\, \sin\phi_2\,,\label{mupairs}
\eea
{\it etc}.  (So $\sum_a \td\mu_a^2=1$.)  
Then, the Ans\"atze (\ref{dgenansatz}) become
\bea
d\hat s^2 &=& \wtd\Delta^{\fft2{D-1}} \, ds_D^2 + \fft1{g^2}\,
\wtd\Delta^{-\fft{D-3}{D-1}}\, \sum_a \wtd X_a^{-1}\, (d\td\mu_a^2
+\td\mu_a^2\, d\phi_a^2)\,,\nn\\
\hat F_{\sst{(D)}} &=& 2g\, \sum_a(\wtd X_a^2\, \td\mu_a^2 - 
\wtd\Delta\, \wtd X_a)\,\ep_{\sst{(D)}}
-\fft1{2g}\, \sum_a \wtd X_a^{-1} \, {* d\wtd X_a}\wedge d(\td\mu_a^2)\,,
\label{dgenansatz2}
\eea
where $\wtd\Delta$ is now given by
\be
\wtd\Delta = \sum_a \wtd X_a\, \td\mu_a^2\,,
\ee
and the index $a$ ranges over $1\le a \le \ft12 N$.  Of course the
quantities $\wtd X_a$ satisfy $\prod_a \wtd X_a=1$, by virtue of
(\ref{prodcon}).  Equations (\ref{dgenansatz2}) are precisely of the
form of the Ans\"atze obtained in \cite{ten} for the cases of the
$S^7$ reduction of $D=11$ supergravity, and the $S^5$ reduction of
type IIB supergravity, where in each case a truncation to the maximal
abelian subgroup of the full gauge group was made.  (Of course here,
where we are considering just the metric and scalar fields in the
lower-dimensional supergravities, the surviving $U(1)^4$ and $U(1)^3$
gauge fields of the maximal abelian truncation are also set to zero.)

     In the case when $N$ is odd, a similar type of specialization
leads us back to a consistent embedding of the form that was derived in
\cite{ten} for the maximal abelian $U(1)^2$ truncation of $D=7$
gauged supergravity into $D=11$ supergravity compactified on $S^4$.
In this case, we make pairwise identifications for the first $(N-1)$ 
$X_i$ and $\mu_i$, as in (\ref{xpairs}) and (\ref{mupairs}), and keep
the single left over quantities 
$\wtd X_0\equiv X_N$, and $\td\mu_0\equiv \mu_N$ as they are.  Now, we
find that the As\"atze (\ref{dgenansatz}) become
\bea
d\hat s^2 &=& \wtd\Delta^{\fft2{D-1}} \, ds_D^2 + \fft1{g^2}\,
\wtd\Delta^{-\fft{D-3}{D-1}}\, \Big(\wtd X_0^{-1}\, d\td\mu_0^2 +
\sum_a \wtd X_a^{-1}\, (d\td\mu_a^2
+\td\mu_a^2\, d\phi_a^2)\Big)\,,\nn\\
\hat F_{\sst{(D)}} &=& 2g\, \sum_a(\wtd X_a^2\, \td\mu_a^2 -  
\wtd\Delta\, \wtd X_a)\,\ep_{\sst{(D)}} + 
g\, \wtd\Delta\, \wtd X_0\, \ep_{\sst{(D)}}\nn\\
&&-\fft1{2g}\, \wtd X_0^{-1}\, {*d\wtd X_0}\wedge d(\td\mu_0^2) 
-\fft1{2g}\, \sum_a \wtd X_a^{-1} \, {* d\wtd X_a}\wedge d(\td\mu_a^2)\,,
\label{dgenansatz3}
\eea
where $\wtd\Delta$ is given by
\be
\wtd\Delta =  \wtd X_0\, \td\mu_0^2 +\sum_a \wtd X_a\, \td\mu_a^2\,,
\ee
and we are taking $a$ to range over $1\le \ft12 (N-1)$ here.  Note that
the tilded fields satisfy the constraint $\wtd X_0 = \prod_a\wtd X_a^{-2}$.

\section{Domain Walls as Distributed $p$-branes}
\label{pBraneOrigin}

    Having obtained the reduction Ansatz for the scalar potentials of
lower-dimensional gauged supergravities from Kaluza-Klein sphere
reductions, it is straightforward to oxidize the AdS domain wall
solutions obtained in section 2.2 to higher dimensions.  In this
section we show that the AdS$_7$, AdS$_4$ and AdS$_5$ domain wall
solutions can be viewed as distributed M5-branes, M2-branes 
and D3-branes respectively.  

\subsection{AdS$_7$ domain walls as distributed M5-branes}

         Using the reduction Ansatz, we find that the oxidation of
the AdS$_7$ domain wall to $D=11$ becomes
\be
d\hat s_{11}^2=H^{-1/3}(-dt^2 + d\vec x\cdot d\vec x) + H^{2/3}\, ds_5^2
\,,\qquad \hat F_\4 ={\hat * (dt\wedge d^5x \wedge dH^{-1})}\,,
\label{m5dist}
\ee
where
\be
H=\fft{1}{g^3\, r^3\, \Delta}\, \qquad 
\Delta = (H_1\cdots H_5)^{1/2}\sum_{i} \fft{\mu_i^2}{H_i}\,.\label{m5hexp}
\ee
The transverse-space metric $ds_5^2$ is given by
\be
ds_5^2 =\fft{\Delta\, dr^2}{\sqrt{H_1\cdots H_5}} + r^2 \sum_i H_i\, d\mu_i^2 
\,,
\ee
which is in fact a flat Euclidean metric $ds_5^2=dy^m dy^m$ after
making the coordinate transformation
\be
y_i = r\sqrt{H_i}\, \mu_i\,.
\ee

         The solution (\ref{m5dist}) describes a continuous
distribution of M5-branes, and the harmonic function $H$ can be
expressed as
\be
H=g^{-3} \int \fft{\sigma(\vec y{\,}')\, d^5y'}{|\vec y-\vec y{\,}'|^3}\,.
\label{m5harm}
\ee
Various cases arise, depending on how many of the $\ell_i$ parameters
are non-zero.  Without loss of generality, we may consider the cases
where the first $n$ parameters $\ell_i^2$ are non-vanishing, for each
$n$ in the range $0\le n\le 4$.  (Note that there is no need to
consider the case with 5 non-vanishing parameters since, as we showed
previously, this is a degenerate situation that is equivalent to having
just 4 non-vanishing parameters.)  We find that the charge
distribution functions $\sigma$ in the various cases are given by
\bea
n=0:
&&\sigma= \delta^\5(\vec y)\,,\nn\\
n=1:
&&\sigma=\fft{1}{2\ell_1}\Theta(1-\fft{y_1^2}{\ell_1^2})\, 
\delta^\4(\vec y_{2,3,4,5})\,,\nn\\
n=2:
&&\sigma=\fft{1}{2\pi \ell_1\ell_2}\, (1 -\fft{y_1^2}{\ell_1^2} -
\fft{y_2^2}{\ell_2^2})^{-1/2}\, \Theta(1 -\fft{y_1^2}{\ell_1^2} -
\fft{y_2^2}{\ell_2^2})\, \delta^\3(\vec y_{3,4,5})\,,\nn\\
n=3:
&&\sigma=\fft{1}{2\pi\ell_1\ell_2\ell_3}
\delta(1-\sum_{i=1}^3\fft{y_i^2}{\ell_i^2})\,
 \delta^\2(\vec y_{4,5})\,,\label{m5distfun}\\
n=4:
&&\sigma= \fft{1}{4\pi^2\ell_1\cdots \ell_4}\Big[
-(1-\sum_{i=1}^4 \fft{y_i^2}{\ell_i^2})^{-3/2}
\Theta(1-\sum_{i=1}^4 \fft{y_i^2}{\ell_i^2})\nn\\
&&\qquad \qquad\qquad + 2(1-\sum_{i=1}^4 \fft{y_i^2}{\ell_i^2})^{-1/2}
\delta(1-\sum_{i=1}^4 \fft{y_i^2}{\ell_i^2})\Big]
\,\delta(y_5)\,.\nn
\eea
The distribution for $n=2$ with $\ell_1=\ell_2$ were also given in
\cite{KLT}.
We may note an interesting relation among these distributions, which
will persist in future sections and was first noticed in \cite{fgpw2}.
If the $n=4$ distribution is integrated along one of its principle
axes (that is, in the direction of one of the $y_i$), and a factor of
$\delta(y_i)$ is then inserted, the result is the $n=3$ distribution.
And if the $n=3$ distribution is integrated similarly, the result is
the $n=2$ distribution---et cetera.  As we explained earlier, there is
{\it a priori} no preferred choice of five-parameter solution: for
example, if all the $\ell_i$ were equal, there would be no preferred
spherically symmetric distribution of branes since all of them lead to
the same perfect $\hbox{AdS}_7 \times S^4$ exterior.  (This is because
the five-parameter solution is degenerate owing to the symmetry of the
solution given by (\ref{ellTransform}).)

However, there is a unique five-parameter solution which shares the
property that integrating it along some principle axis leads to the
$n=4$ solution.  It is 
\be
  n=5: \ \ \ 
  \sigma=\fft{1}{2\pi^2\ell_1\ell_2\ell_3\ell_4\ell_5}
  \delta'(1-\sum_{i=1}^5\fft{y_i^2}{\ell_i^2})\,.\label{M5Gen}
\ee
 All the other distributions, (\ref{m5distfun}), can be regarded as
limits of (\ref{M5Gen}) where some of the $\ell_i$ are taken to $0$.
Such a limiting process is equivalent to integrating along the
corresponding $y_i$ directions and then inserting factors of
$\delta(y_i)$.

         The above discussion is applicable for positive values of
$\ell_i^2$.  The solutions with negative $\ell_i^2$ can be mapped to
the solutions with positive $\ell_i^2$ using the transformation
(\ref{ellTransform}), as discussed in section 2.2.

\subsection{AdS$_4$ domain walls as distributed M2-branes}

         Analogously, we find that the oxidation of
the AdS$_4$ domain wall to $D=11$ becomes
\be
d\hat s_{11}^2=H^{-2/3}(-dt^2 + d\vec x\cdot d\vec x) + H^{1/3}\, ds_8^2
\,,\qquad \hat F_\4 =dt\wedge d^2x \wedge dH^{-1}\,,
\label{m2dist}
\ee
where
\be
H=\fft{1}{g^6\, r^6\, \Delta}\, \qquad 
\Delta = (H_1\cdots H_8)^{1/2}\sum_{i=1}^8 \fft{\mu_i^2}{H_i}\,.
\ee
The transverse-space metric $ds_8^2$ is given by
\be
ds_8^2 =\fft{\Delta\, dr^2}{\sqrt{H_1\cdots H_8}} + r^2 \sum_i H_i\, d\mu_i^2 
\,,
\ee
which is in fact a flat Euclidean 8-metric $ds_8^2=dy^m dy^m$ after
making the coordinate transformation
\be
y_i = r\sqrt{H_i}\, \mu_i\,.
\ee

         The solution (\ref{m2dist}) describes a continuous
distribution of M2-branes, and the harmonic function $H$ can be
expressed as
\be
H=g^{-6} \int \fft{\sigma(\vec y{\,}')\, d^8y'}{|\vec y-\vec y{\,}'|^6}\,.
\ee
For the various cases with $0\le n\le 7$ parameters $\ell_i$, we find
that the charge distribution functions $\sigma$ are given by
\bea
n=0:
&&\sigma= \delta^\8(\vec y)\,,\nn\\
n=1:
&&\sigma=\fft{8}{3\pi\ell_1}\, (1-\fft{y_1^2}{\ell_1^2})^{3/2}\, 
\Theta(1-\fft{y_1^2}{\ell_1^2})\, \delta^\7(\vec y_{2,3,4,5,6,7,8})
\,,\nn\\ 
n=2:
&&\sigma =\fft{2}{\pi\ell_1\ell_2}\, (1-\fft{y_1^2}{\ell_1^2}
-\fft{y_2^2}{\ell_2^2})\, \Theta(1-\fft{y_1^2}{\ell_1^2}
-\fft{y_2^2}{\ell_2^2})\, \delta^\6(\vec y_{3,4,5,6,7,8})
\,,\nn\\
n=3:
&&\sigma=\fft{4}{\pi^2\ell_1\ell_2\ell_3}\, 
(1 -\sum_{i=1}^3\fft{y_i^2}{\ell_i^2})^{1/2}\, 
\Theta(1 -\sum_{i=1}^3\fft{y_i^2}{\ell_i^2})\, 
\delta^\5(\vec y_{4,5,6,7,8})\,,\nn\\
n=4:
&&\sigma=\fft{2}{\pi^2\ell_1\cdots\ell_4}\, 
\Theta(1 -\sum_{i=1}^4\fft{y_i^2}{\ell_i^2})\, 
\delta^\4(\vec y_{5,6,7,8})\,,\label{m2distfun}\\
n=5:
&&\sigma=\fft{2}{\pi^3\ell_1\cdots\ell_5}\, 
(1 -\sum_{i=1}^5\fft{y_i^2}{\ell_i^2})^{-1/2}\, 
\Theta(1 -\sum_{i=1}^5\fft{y_i^2}{\ell_i^2})\, 
\delta^\3(\vec y_{6,7,8})\,,\nn\\
n=6:
&&\sigma=\fft{2}{\pi^3\ell_1\cdots\ell_6}\,
\delta(1 -\sum_{i=1}^6\fft{y_i^2}{\ell_i^2})\,
\delta^\2(\vec y_{7,8})\,,\nn\\
n=7:
&&\sigma=\fft{1}{\pi^4\ell_1\cdots\ell_7}\, 
\Big[-(1 -\sum_{i=1}^7\fft{y_i^2}{\ell_i^2})^{-3/2}\,
\Theta(1 -\sum_{i=1}^7\fft{y_i^2}{\ell_i^2})\nn\\
&&\qquad\qquad\qquad  +2(1 -\sum_{i=1}^7\fft{y_i^2}{\ell_i^2})^{-1/2}\,
\delta(1 -\sum_{i=1}^7\fft{y_i^2}{\ell_i^2})\Big]\,
\delta(y_8)\,.\nn
\eea
The distribution for $n=2$ with $\ell_1=\ell_2$ was also
given in \cite{KLT}.
Note that the 8-parameter case is degenerate, since it is equivalent
to a 7-parameter solution.  However, as in the case of
M5-branes, the distribution
\be
 n=8: \ \ \ 
 \sigma=\fft{2}{\pi^4\ell_1\cdots\ell_8}\, 
  \delta'(1 -\sum_{i=1}^8\fft{y_i^2}{\ell_i^2})\label{M2Gen}
\ee
 can be regarded as the ``parent'' of all the others in the sense that
all the distributions in (\ref{m2distfun}) are limits of (\ref{M2Gen})
where some of the $\ell_i$ are taken to $0$.

         The discussion is applicable for positive values of
$\ell_i^2$.  The solutions with negative $\ell_i^2$ can be mapped to
the solutions with positive $\ell_i^2$ using the transformation
(\ref{ellTransform}), as discussed in section 2.2.

\subsection{AdS$_5$ domain walls as distributed D3-branes}

     Analogously, we find that the oxidation of the AdS$_5$ domain
wall to $D=10$ becomes
\be
d\hat s_{10}^2=H^{-1/2}(-dt^2 + d\vec x\cdot d\vec x) + H^{1/2}\, ds_6^2
\,,\qquad \hat G_\5 =dt\wedge d^3x \wedge dH^{-1}\,,
\label{d3dist}
\ee
where
\be
H=\fft{1}{g^4\, r^4\, \Delta}\, \qquad 
\Delta = (H_1\cdots H_6)^{1/2}\sum_{i=1}^6 \fft{\mu_i^2}{H_i}\,.
\ee
The transverse-space metric $ds_6^2$ is given by
\be
ds_6^2 =\fft{\Delta\, dr^2}{\sqrt{H_1\cdots H_6}} + r^2 \sum_i H_i\, d\mu_i^2 
\,,
\ee
which is in fact a flat Euclidean 6-metric $ds_6^2=dy^m dy^m$ after
making the coordinate transformation
\be
y_i = r\sqrt{H_i} \, \mu_i\,.
\ee

         The solution (\ref{d3dist}) describes a continuous
distribution of D3-branes, and the harmonic function $H$ can be
expressed as
\be
H=g^{-4} \int \fft{\sigma(\vec y{\,}')\, d^8y'}{|\vec y-\vec y{\,}'|^4}\,.
\ee
For the various $0\le n\le 4$ parameter configurations, we find that
the charge distribution functions $\sigma$ are given by
\bea
n=0:
&&\sigma= \delta^\6(\vec y)\,,\nn\\
n=1:
&&\sigma =\fft{2}{\pi\ell_1} (1-\fft{y_1^2}{\ell_1^2})^{1/2}\,
\Theta(1-\fft{y_1^2}{\ell_1^2})\, \delta^\5(\vec y_{2,3,4,5,6})
\,,\nn\\
n=2:
&&\sigma =\fft{1}{\pi\ell_1\ell_2}\, 
\Theta(1-\fft{y_1^2}{\ell_1^2} -\fft{y_2^2}{\ell_2^2})\,
\delta^\4(\vec y_{3,4,5,6})\,,\nn\\
n=3:
&&\sigma =\fft{1}{\pi^2\ell_1\ell_2\ell_3}\, 
(1-\sum_{i=1}^3\fft{y_i^2}{\ell_i^2})^{-1/2}\, 
\Theta(1-\sum_{i=1}^3\fft{y_i^2}{\ell_i^2})\,
\delta^\3(\vec y_{4,5,6})\,,\label{d3distfun}\\
n=4:
&&\sigma =\fft{1}{\pi^2\ell_1\cdots\ell_4}\, 
\delta(1-\sum_{i=1}^4\fft{y_i^2}{\ell_i^2})\,
\delta^\2(\vec y_{5,6})\,,\nn\\
n=5:
&&\sigma =\fft{1}{2\pi^3\ell_1\cdots\ell_5}\Big[-
(1-\sum_{i=1}^5\fft{y_i^2}{\ell_i^2})^{-3/2}\,
\Theta(1-\sum_{i=1}^5\fft{y_i^2}{\ell_i^2})\nn\\
&&\qquad\qquad\qquad  +2(1-\sum_{i=1}^5\fft{y_i^2}{\ell_i^2})^{-1/2}
\delta(1-\sum_{i=1}^5\fft{y_i^2}{\ell_i^2})\Big]\,
\delta(y_6)\,.\nn
\eea
Various special cases of the above list can also be found
in \cite{KLT,fgpw2}.
Note that the 6-parameter case is degenerate, being equivalent to a
5-parameter solution through the transformation (\ref{ellTransform}).  
Once again, though, there is a unique 6-parameter distribution,
  \be
    n=6: \ \ \ 
     \sigma =\fft{1}{\pi^3\ell_1\cdots\ell_6}
      \delta'(1-\sum_{i=1}^6\fft{y_i^2}{\ell_i^2}) \label{D3Gen}
  \ee
such that all the other distributions (\ref{d3distfun}) are limits of
(\ref{D3Gen}) in which some of the $\ell_i$ are taken to $0$.

         The above discussions are applicable for positive values of
$\ell_i^2$.  The solutions with negative $\ell_i^2$ can be mapped to
the solutions with positive $\ell_i^2$ using the transformation
(\ref{ellTransform}), as discussed in section~2.2.

         Note that the charge distributions $\sigma(\vec y)$ have the
same functional form for M2-branes, M5-branes and D3-branes with the 
same $N-n$.

\section{Six-dimensional Gauged Supergravity}
\label{AdS6}

   In our discussion so far, we have focussed on the class of scalar
plus gravity theories that correspond to certain consistent
truncations of the gauged maximal supergravities in $D=4$, 5 and 7.
There is one other case that we wish to consider here, namely gauged
supergravity in $D=6$.  This is an unusual case in that there
apparently does not exist a gauged supergravity with the maximum
number of supersymmetries that are allowed in six dimensions.  The
known $SU(2)$ gauged theory has $N=2$ rather than $N=4$ supersymmetry
\cite{romans}.  It was suggested in \cite{ferrara} that it could be
related to the ten-dimensional massive type IIA theory \cite{roman10}.
It has recently been shown that it can be obtained from an $S^4$
reduction of massive type IIA supergravity \cite{d6gauge}.  Since it
does not have maximal supersymmetry there is nothing to preclude
coupling it to $N=2$ matter.  The six-dimensional scalar plus gravity
theory that we shall consider here can be interpreted as a consistent
truncation of the $SU(2)$ gauged $N=2$ supergravity in $D=6$ coupled
to certain matter multiplets.  (Domain wall solutions in pure $D=6$
gauged supergravity were previously studied in
\cite{dilatonic,skto}.

   The only part of the massive IIA Lagrangian that is
relevant for our truncation comprises the metric, the dilaton and the
4-form field strength:
\be
{\cal L}_{10} = \hat R{*\oneone} -\ft12 {* d\hat\phi}
\wedge d\hat \phi -\ft12 e^{\ft12\hat \phi}\,{*\hat F_\4}
\wedge \hat F_\4 -\ft12 m^2 e^{\ft52 \hat \phi}\, {*\oneone}
\,.\label{d10lag}
\ee
This admits a solution describing an intersection of
D4-branes and D8-branes, namely \cite{youm}
\bea
ds_{10}^2 &=& (g\,z)^{\fft1{12}}\Big( H^{-3/8}\, (-dt^2 + d\vec x\cdot
d\vec x) + H^{5/8}\, (d\vec y^2 + dz^2)\Big)\,,\nn\\
e^{\hat\phi} &=& (g\, z)^{-\fft56} H^{-1/4}\, \qquad
F_4=e^{-\fft12\hat \phi}\, {*(d^5x\wedge dH^{-1})}\,,
\eea
where $g= 3m/\sqrt2$ and the ``harmonic'' function satisfies
\be
z^{-1/3}\, \del_z(z^{1/3}\del_z H) + \square_y H=0\ .\label{d6eom}
\ee
One class of solutions of this differential equation can be expressed as
follows \cite{youm}:
\be
H = c +g^{-10/3}\,
\int \fft{\sigma(\vec y{\,}') d^4y'}{(|\vec y - \vec y{\,}'|^2 + z^2)^{5/3}}
\,,\label{d6harm}
\ee
where $c$ is an integration constant, which we shall set to zero.  It
should be emphasized, however, that this is not the most general
solution to equation (\ref{d6eom}).   If the distribution function $\sigma$
is a delta function, {\it i.e.}\ $\sigma=\delta^\4(\vec y)$, the
metric becomes warped AdS$_6\times S^4$ \cite{yaron}.

  Inspired by the previous examples of M-branes and the D3-brane, let us
consider the following charge distribution, 
\bea
\sigma &=&\fft{2}{9\pi^2\, \ell_1\cdots\ell_4}\Big[
-(1-\sum_{i=1}^4\fft{y_i^2}{\ell_i^2})^{-4/3}\,
\Theta(1-\sum_{i=1}^4\fft{y_i^2}{\ell_i^2}) \nn\\
&&\qquad\qquad\qquad + 3(1-\sum_{i=1}^4\fft{y_i^2}{\ell_i^2})^{-1/3}\,
\delta(1-\sum_{i=1}^4\fft{y_i^2}{\ell_i^2})\Big]\,.
\label{d6n4dis}
\eea
To evaluate $H$ with this distribution, we perform the following
coordinate transformation:
\be
y_i = \sqrt{r^2 + \ell_i^2}\, \mu_i\,,\qquad
z=\sqrt{r^2 + \ell_0^2}\, \mu_0\,,\label{d4d8redef}
\ee
where $\sum_\a\mu_\a^2=1$, summing over $\a=(0,i)$.  Note, however,
that although we included a parameter $\ell_0$ for the sake of
symmetry, it should actually be set to zero.  Next, we define $H_\a =
1 + \ell_\a^2/r^2$ (and so $H_0=1$).  It then follows that the
function $H$ becomes
\be
H=\fft1{(g\,r)^{10/3}\, \Delta}\,,\qquad
\Delta =(\prod_\a H_\a)^{1/2}\, \sum_\a \fft{\mu_\a^2}{H_\a}\,.
\label{d48hexp}
\ee
In terms of the redefined coordinates $y^i$ and $z$, the
transverse-space metric is
\be
dy^i dy^i + dz^2 = (\prod_\a H_\a)^{-1/2}\, \Delta\, dr^2 + r^2\, H_\a\,
d\mu_\a^2\,.
\ee
Thus the ten-dimensional metric of the D4-D8-brane becomes
\be
ds_{10}^2 = \mu_0^{1/12}\Big[ (gr)^{4/3}\, \Delta^{3/8}\,
dx^\mu dx_\mu + \Delta^{3/8} (\prod_\a H_\a)^{-1/2}\,
\fft{dr^2}{g^2 r^2} + \Delta^{-5/8}\, g^{-2}
\sum_\a H_\a d\mu_\a^2\Big]\,.
\ee
If we now define
\be
\wtd \Delta= (\prod_\a H_\a)^{-5/16} \Delta = (\prod_\a H_\a)^{3/16}
\sum \fft{\mu_\a^2}{H_\a} \equiv \sum_\a X_\a\, \mu_\a^2\,,
\ee
the ten-dimensional D4-D8-brane solution can be expressed in the
following abstract form:
\bea
ds_{10}^2 &=& \mu_0^{1/12}(\prod_\a X_\a)^{1/8}\Big(
\wtd\Delta^{3/8}\, ds_6^2 + g^{-2}\, \wtd \Delta^{-5/8}\,
\sum_\a X_\a^{-1}\, d\mu_\a^2\Big)\,,\nn\\
e^{\ft12\hat \phi}{*\hat F_\4} &=&
g \sum_\a (2X_\a^2\, \mu_\a^2 -\wtd \Delta\, X_\a)
\epsilon_\6 -\ft13 g \wtd \Delta\, X_0 \epsilon_\6
+\fft1{2g}\sum_\a X_\a^{-1} { *dX_\a}\, d(\mu_\a^2)\,,\nn\\
e^{\hat \phi} &=& \mu_0^{-5/16}\, \wtd \Delta^{1/4}\,
(\prod_\a X_\a)^{-5/4}\,,\label{ads6red}
\eea
where the six-dimensional metric and scalars $X_\a$  are given by
\bea
ds_6^2 &=& (gr)^{4/3} (H_1\cdots H_4)^{1/8}\, dx^\mu dx_\mu
+(H_1\cdots H_4)^{-3/8} \fft{dr^2}{g^2r^2}\,,\nn\\
X_i &=& (H_1\cdots H_4)^{3/16} H_i^{-1}\,, \qquad
X_0=(H_1\cdots H_4)^{3/16}\,.\label{ads6sol}
\eea
Note that the quantities $X_\a$ are not independent, but are 
subject to the constraint
\be
X_0=(X_1\cdots X_4)^{-3/4}\,.\label{ads6cons}
\ee

      Having written the solutions in an abstract form, we can now
propose to view (\ref{ads6red}) as a general Kaluza-Klein reduction
Ansatz, for arbitrary six-dimensional fields, whose equations of
motion can be derived by substituting the Ansatz into the
ten-dimensional equations of motion for the massive IIA theory.  (Of
course, for this to succeed, it is necessary that the reduction Ansatz
(\ref{ads6red}) be a consistent one, which appears to be the case.)
In particular, we can determine the structure of the scalar potential.
This is described by four independent scalars, parameterised by
$X_\a$, subject to the constraint (\ref{ads6cons}).

\subsection{$D=6$ potential}

   We can work this out by starting from the Ansatz (\ref{ads6red}) for
the 4-form field strength.  Its Bianchi identity gives us the
equations of motion for the $X_\a$.  Thus we can immediately read off
that
\be
\square \log X_\a = 4 g^2\, X_\a^2 -2 g^2\, X_\a\, \sum_\b X_\b -
\ft23 g^2\, X_0\, X_\a + Q\,,\label{d6eom1}
\ee
where $Q$ is an undetermined quantity independent of $\a$.  (The
ambiguity here arises, as usual, because $\sum_\a d(\mu_\a^2) = 0$, by
virtue of $\sum_\a \mu_\a^2 =1$.)   We can determine $Q$ by noting
that $X_0= (\prod_i X_i)^{-3/4}$, and hence
\be
\ft43 \log X_0 + \sum_i \log X_i=0\,.
\ee
(Recall we are splitting the indices as $\a=(0,i)$.)  Thus if we take
this particular sum over terms in (\ref{d6eom1}), the left-hand side
must vanish, hence giving us an equation for $Q$:
\be
Q= g^2\, X_0\, \sum_i X_i - \ft13 g^2\, X_0^2 - \ft34 g^2\, \sum_i
X_i^2 + \ft38 g^2\, (\sum_i X_i)^2\,.
\ee
In particular, this means, plugging back into (\ref{d6eom1}), that we
have
\bea
\square \log X_0 &=& g^2\, X_0^2 - g^2 X_0\, \sum_i X_i - \ft34 g^2\,
\sum_i X_i^2 + \ft38 g^2\, (\sum_i X_i)^2\,,\nn\\
\square\log X_i &=& 4 g^2\, X_i^2 - 2g^2\, X_i\, \sum_j X_j -\ft83
g^2\, X_0\, X_i  \label{d6eom2}\\
&&+ \ft38 g^2\, (\sum_j X_j)^2 -\ft34 g^2\, \sum_j X_j^2 + g^2\, X_0\,
\sum_j X_j -\ft13 g^2\, X_0^2\,.\nn
\eea

   We now look for a 6-dimensional potential $V$ such that
\be
e^{-1}\, {\cal L}_6 =R - \ft12(\del\vec\varphi)^2 - V
\label{d6lag}
\ee
reproduces the above equations of motion, where $\vec\varphi$ is a
4-vector of dilatons.  We express $X_0$ and $X_i$ in terms of these in
the usual way, $X_\a = e^{-\fft12\vec b_\a\cdot\vec\varphi}$.  At this
stage, let us just present the results.  We find that to reproduce the
equations of motion (\ref{d6eom2}) correctly, we should take the
vectors $\vec b_\a$ to satisfy the dot products
\be
\vec b_i\cdot\vec b_j = 8\delta_{ij} - \ft32\,,\qquad \vec b_0\cdot
\vec b_i = -\ft32\,,\qquad \vec b_0\cdot\vec b_0 = \ft92\,.
\ee
Of course the identity $X_0= (\prod_i X_i)^{-3/4}$ implies that we
will also have
\be
\vec b_0 + \ft34\sum_i \vec b_i = 0\,.
\ee

    The required scalar potential then turns out to be
\be
V= -\ft12 g^2\, \Big( (\sum_i X_i)^2 - 2 \sum_i X_i^2
+ \ft83 X_0\, \sum_i X_i - \ft89 X_0^2 \Big)\,.\label{d6pot}
\ee
Note that if we specialise to the case where $X_1=X_2\equiv\wtd X_1$,
$X_3=X_4\equiv\wtd X_2$, $X_0\equiv\wtd X_0$, we obtain the 
potential
\be
V= \ft49 g^2\, (\wtd X_0^2 - 9 \wtd X_1\, \wtd X_2 - 6 \wtd X_0\, \wtd
X_1 -6 \wtd X_0\, \wtd X_2)\,.
\ee
Note that a further specialisation $\wtd X_1=\wtd X_2$ gives back the
potential of $D=6$ gauged supergravity, obtained from massive type IIA
in \cite{d6gauge}.

\subsection{AdS$_6$ domain walls as distributed D4-D8-branes}

         The scalar Lagrangian (\ref{d6lag}) with potential
(\ref{d6pot}) admits a six-dimensional domain wall solution with four
parameters $\ell_i$, given by (\ref{ads6sol}).  From the $D=10$ standpoint,
the solution can be viewed as a distribution of D4-D8-branes with
charge distribution function (\ref{d6n4dis}).  For the various numbers of
non-vanishing parameters $\ell_i$, the charge distribution function $\sigma$
is given by
\bea
n=0:&& \sigma =\delta^\4(\vec y)\,,\nn\\
n=1:&& \sigma =\fft{\Gamma(5/3)}{\sqrt\pi\,\Gamma(7/6)\,\ell_1}\,
(1-\fft{y_1^2}{\ell_1^2})^{1/6}\,
\Theta(1-\fft{y_1^2}{\ell_1^2})\, \delta^\3(\vec y_{2,3,4})\,,\nn\\
n=2:&& \fft{2}{3\pi\,\ell_1\ell_2}\,
(1-\fft{y_1^2}{\ell_1^2} -\fft{y_2^2}{\ell_2^2})^{-1/3}\,
\Theta(1-\fft{y_1^2}{\ell_1^2} -\fft{y_2^2}{\ell_2^2})\,
\delta^\2(\vec y_{3,4})\,,\label{d48distfun}\\
n=3:&& \fft{\Gamma(5/3)}{\pi^{3/2}\,\Gamma(1/6)\,\ell_1\ell_2\ell_3}\,
(1-\sum_{i=1}^3\fft{y_i^2}{\ell_i^2})^{-5/6}\,
\Theta(1-\sum_{i=1}^3\fft{y_i^2}{\ell_i^2})\, \delta(y_4)\,,\nn\\
n=4:&&\sigma =\fft{2}{9\pi^2\, \ell_1\cdots\ell_4}\Big[
 -(1-\sum_{i=1}^4\fft{y_i^2}{\ell_i^2})^{-4/3}\,
\Theta(1-\sum_{i=1}^4\fft{y_i^2}{\ell_i^2}) \nn\\
&&\qquad\qquad\qquad +3(1-\sum_{i=1}^4\fft{y_i^2}{\ell_i^2})^{-1/3}\,
\delta(1-\sum_{i=1}^4\fft{y_i^2}{\ell_i^2})\Big]\,.\nn
\eea

    A remark about the evaluation of the integral (\ref{d6harm}) with
the charge distributions (\ref{d48distfun}) is in order here.  In the
previous cases, of the D3-brane, M2-brane and M5-brane, the analogous
integrals could be studied rather easily (as in \cite{KLT}) 
by first specialising the
coordinates $\vec y$ in the resulting harmonic functions $H(\vec y)$
to simple values for which the integration is more easily performed.
$H(\vec y)$ for this specialised range of $\vec y$ coordinate values
can then be matched to the general solution of Laplace's equation,
expressed as a sum over a complete set of elementary solutions,
yielding a unique determination of the expansion coefficients.  This
then establishes that provided $H(\vec y)$ evaluated over the
specialised coordinate range agrees with the claimed general form for
$H$ evaluated over the same range, then the integral over the
distribution must be in agreement with the claimed expression for
$H(\vec y)$ for arbitrary coordinate values $\vec y$.  Thus, for
example, when evaluating (\ref{m5harm}) for the
$\ell_i=(\ell_1,\ell_2,0,0,0)$ charge distribution given in
(\ref{m5distfun}), it suffices to evaluate the integral for $H(\vec
y)$ at $y_1=y_2=0$, and verify that it agrees with $H$ given by
(\ref{m5hexp}) in this restricted coordinate region.  One can, of
course, instead directly evaluate the integral (\ref{m5harm}) for
$H(\vec y)$ with generic values for $\vec y$, but this is quite a bit
more complicated in practice.  In our present case, with the
evaluation of (\ref{d6harm}) for the D4/D8-brane distributions
(\ref{d48distfun}), one is on less solid ground if one tries to invoke
uniqueness to argue that it suffices to evaluate the integral for
$H(\vec y)$ at special values of $\vec y$, since, as we remarked,
(\ref{d6harm}) is not the most general solution of the equations for
D4/D8-brane intersections.  We have therefore performed explicit
integrations for generic $\vec y$ coordinate values, in order to
verify that the distribution integrals do indeed correctly reproduce
the functions given by (\ref{d48hexp}).  

         The discussion is applicable only for non-negative
$\ell_i^2$.  For solutions with negative $\ell_i^2$, it can no longer
be viewed as distributions of solutions of the form (\ref{d6harm}).
This is not surprising, since unlike the D3-brane and M-branes, the
harmonic function here (\ref{d6harm}) is not the most general solution
of its equation (\ref{d6eom}).  Thus whilst for the previous cases
the solutions have to be expressible in terms of distributions of
D3-branes or M-branes, here it is more of a ``miracle'' that the
six-dimensional AdS domain walls with positive $\ell_i^2$ can be
expressed as distributions of D4-D8-branes, with the form
(\ref{d6harm}).

    It is worth remarking that the D4-D8-brane system also has other
features that are rather different from those in the
D3-brane, M2-brane and M5-brane systems.  The distributed D3-branes,
M2-branes and M5-branes in $D=10$ or $D=11$ admit lower-dimensional
interpretations as extremal
solutions in the gauged maximal supergravities in $D=5$, $D=4$ and
$D=7$ respectively.  By contrast, the D4-D8-brane systems admit a
lower-dimensional interpretation in $D=6$, where there apparently does
not exist any maximally-supersymmetric gauged supergravity.  There
does exist a gauged theory with one half of the maximal supersymmetry
in $D=6$ \cite{romans} (namely $N=2$), and an $SU(2)$ gauge group, 
which arises as a
consistent $S^4$ reduction of massive type IIA supergravity \cite{d6gauge}.
However, this theory contains only a single scalar field $X$, which
corresponds to the truncation $X_1=X_2=X_3=X_4\equiv X$, $X_0=X^{-3}$ 
in the six-dimensional potential (\ref{d6pot}).  Thus all four of the
original $\ell_i$ parameters in the solutions (\ref{ads6sol}) must be
set equal in order that the configuration be a solution purely within
the $SU(2)$-gauged $N=2$ supersymmetric theory.
We have shown that the more general
configurations (\ref{ads6sol}) with all four $\ell_i$ different are solutions
of the six-dimensional theory (\ref{d6lag}) and (\ref{d6pot}), which
can also be obtained as a consistent dimensional reduction of the
massive type IIA supergravity.  However, this six-dimensional theory
is not a truncation of the pure gauged supergravity in $D=6$; rather,
it is a truncation of a gauged supergravity coupled to
matter.\footnote{A closely parallel situation which, however, highlights
the distinction is that of the two-charge AdS black hole solutions of
maximal ($N=2$) $SO(5)$-gauged supergravity in $D=7$.  If the charges
are set equal, the black holes can be viewed as solutions in pure
$N=1$ $SU(2)$-gauged supergravity, but if the charges are unequal,
then from an $N=1$ standpoint the black holes are solutions of $N=1$
gauged supergravity coupled to the matter that results from viewing
the $N=2$ pure supergravity as an $N=1$ theory.  The difference in
$D=6$ is that there {\it is} no maximal gauged supergravity that could
provide the alternative ``$N=2$ viewpoint.''}

\section{Analysis of the spectrum}
\label{Spectrum}

The correlations functions (and the spectrum) of the corresponding
strongly coupled gauge theory can be analyzed by studying the wave
equations in these gravitational backgrounds.  In particular, the
simplest two-point function is that of the operator ${\cal O} \sim
{\rm Tr} F^2+\cdots$ which couples to the $s$-wave dilaton $\phi$.  In
ten dimensions the dilaton 
obeys a free wave equation. In
lower dimensions, a consistent truncation ensures that the dilaton can be
taken to be independent of the spherical coordinates, and the $s$-wave
dilaton does not participate in the gauged supergravity scalar potential.
Generically we expect there always to be a minimal scalar in our
asymptotically AdS geometries: a field $\phi$ which obeys the corresponding
Laplace equation,
\begin{equation}
\partial_\mu(\sqrt{-g}g^{\mu\nu}\,\partial_\nu
\phi)=0\,.\label{wave}
\end{equation}
We choose the Ansatz $\phi=e^{i\omega t}\, \chi(r)$, where $\omega$
characterizes the energy level of the solution, specified by
$\chi(r)$, which is chosen to depend only on the radial coordinate
$r$. Then for the distributed D3-branes and M-branes discussed in the
previous sections, the wave equation (\ref{wave}) takes the following
form:\footnote{The glue-ball ($0^{++}$) spectra for the Euclidean
spinning brane backgrounds are determined by the same wave equations
and thus the same analysis is applicable in these cases also. (See
\cite{CG2} and references therein.)}
\begin{eqnarray}
{r}^{-1}\partial_r\left[ r^{-1}\prod_{i=1}^{N}
\sqrt{r^2+\ell_i^2}\, \partial_r\chi \right]={\cal Q}\chi
\,,\label{wave1}
\eea
where ${\cal Q} = -\omega^2/g^{N/2}$ 
and $N=8$, 6 and 5, for the M2-brane, D3-brane
and M5-brane cases, respectively.  The wave equation for the
distributed D4-D8-brane cases has a somewhat different form, given by
\begin{eqnarray}
{r}^{-1}\partial_r\left[ r^{1/3}\prod_{i=1}^{4}
\sqrt{r^2+\ell_i^2}\, \partial_r\chi \right]={\cal Q}\chi
\,.\label{wave2}
\eea
Here ${\cal Q}=-\omega^2/g^{10/3}$.

\subsection{Spectrum with ``Euclidean''  parameters}

In the approach to modelling confining gauge theories through higher
dimensional branes at finite temperature (initiated in \cite{witHol2}), the
spectrum of low-energy excitations in the gauge theory is the same as the
spectrum of supergravity modes in the dual asymptotically anti-de Sitter
geometry.  By solving the eigenvalue problems (\ref{wave1}) and
(\ref{wave2}) rotated to Euclidean signature we are gleaning information
regarding the dual gauge theory.  Although it has been argued in
\cite{cort} that some of these gauge theories are indeed confining, the
analysis of \cite{CG2} suggests that often they are not.  The issue raised
in \cite{CG2} is that the ratio of the tree-level thermal mass to the
temperature vanishes for some of the gauginos in a limit corresponding to
one of the domain walls we have considered in this paper.  The would-be
confining gauge theory approaches a non-confining supersymmetric theory in
the limit.  What sets the mass scale of the spectrum in such a case is not
the scale of confinement, but rather the Higgs VEV that determines the
state on the Coulomb branch.  We will see from supergravity calculations
that the spectrum can be discrete, or a continuum above a gap, or a
continuum with no gap.  The field theory understanding of these results is
not very satisfactory, although some attempts at explanation have been made
in \cite{fgpw2,BrandhuberSfetsos,ChepelevRoiban,GiddingsRoss} (mainly in
the Lorentzian cases).

When Wick-rotating (\ref{wave1}) and (\ref{wave2}) to Euclidean signature,
we have a choice of whether to change the signs of the quantities
$\ell_i^2$.  The relationship of some of these geometries to spinning brane
geometries \cite{MyersChamblin,CG1} suggests that the natural choice is to
send $\ell_i^2 \to -\ell_i^2$ in the Wick rotation.  The reason is that the
$\ell_i$ are proportional to the angular momenta in the spinning brane
geometries, and angular momentum becomes imaginary on Wick rotation.  Thus
we will refer to $\ell_i^2 < 0$ as a choice of Euclidean parameters.  For
M2-branes, D3-branes, and M5-branes, the transformation
(\ref{ellTransform}) allows us to map Euclidean parameters back to
Lorentzian parameters.  The only Lorentzian cases that need to be addressed
separately are for the D4-D8-brane system; see section~\ref{Lorentz}.

 
 The wave equations (\ref{wave1}, \ref{wave2}) can be cast in the
following form:
\begin{equation}
\left[ \partial_y f(y) \partial_y - {\cal Q} \right] 
\chi = 0\,, \label{wavee}
\end{equation}
where $y\equiv r^2-r_H^2$ and $r_H^2$ is an adjustable parameter.  For
M2-branes, D3-branes, and M5-branes, we have 
  $$
   f(y) = \sqrt{\prod_{i=1}^N (y+r_H^2-\ell_i^2)}\,,
  $$
 where as usual $N=8$, $6$, and $5$ in the respective cases.  Let us choose
$r_H^2$ so that the most positive root of $f(y)$ is at $y=0$.  We will
solve the equation (\ref{wavee}) subject to regular boundary conditions at
$y=0$ and $y=\infty$, which is the boundary of AdS.  We are thus excluding
from consideration geometries like spherical shells of branes: the shell
would sit at a finite value of $y$ where the norm of the Killing vector
$\partial/\partial t$ is finite.  To put it another way, we set one of the
$\ell_i$ to $0$ so that we have only $N-1$ free parameters.  Generically,
(\ref{wavee}) cannot be solved exactly (although in special cases it
reduces to a form of the hypergeometric equation
\cite{fgpw2,BrandhuberSfetsos}).  However we may straightforwardly extract
the qualitative features of the spectrum by analyzing the asymptotics of
(\ref{wavee}) near the horizon ($y \to 0$) and near the boundary of AdS ($y
\to \infty$).  The qualitative features depend only on how many of the
parameters $\ell_i$ are nonzero.  Let us take $\ell_i^2 = -1$ for $n_e$
values of $i$, and set $\ell_i^2 = 0$ for the other $N-n_e$ values.  Then
the function $f(y)$ takes the form:
\begin{equation}
 f(y)=4 y^{\fft12 n_e}\,  (y+1)^{\fft12(N-n_e)}\ ,
\end{equation}
with $N=8$, 6, $\ft{16}{3}$ and 5 for the M2-brane, D3-brane, D4-D8-brane and
M5-brane cases respectively. Note that $f(y)$ is an increasing
positive function of $y$ with the asymptotic behavior
\bea
f(y) &\sim& y^{\fft12 N} \qquad \hbox{for $y \to \infty$}\ ,\nn\\
f(y) &\sim& y^{\fft12 n_e} \qquad \hbox{for $y \to 0$}\,.
\eea

   For the analysis of the spectrum it is useful to cast the wave
equations (\ref{wavee}) into the form of the Schr\" odinger equation.  This
is accomplished by introducing the new coordinate $z$, defined by
$\partial y/\partial z = \sqrt{f(y)}$, and setting $\chi = \psi\,
f^{-1/4}$.  We then find that (\ref{wavee}) indeed takes the form of
Schr\" odinger equation:
\bea
&&\left[ -\partial_z^2 + V(z) \right] \psi = -{\cal Q} \psi\ ,\nn\\
&&V = {1 \over 4} \partial_z^2 \log f + {1 \over 16}
    (\partial_z \log f)^2 \,.
\eea
In order to determine the  qualitative behavior of the spectrum, it now 
suffices to analyze the ``Schr\"odinger potential'' $V(z)$ near the
endpoints.

The coordinate $z$ was chosen as an increasing function of $y$. The
potential has the following properties.   Firstly, at the AdS
boundary, which is reached as $y\to \infty$,  we have $z\to z'$, where
$z'$ is finite, and the potential blows up quadratically:
\be
 {\rm AdS}\ \ {\rm boundary}:\ \ V={{C_N}\over
{(z-z')^2}} \, \label{CN}\,.
\ee
Here, the coefficient $C_N$ depends on $N$ (but not on $n_e$).  At the
``horizon'' boundary $y\to 0$, we have $z\to {\tilde z}$ with ${\tilde
z}$ finite for $n_e=1,2,3$, whilst ${\tilde z}\to -\infty$ for
$n_e=4,5,6,7$.  At this boundary the potential has the form
\bea
{\rm Horizon\  boundary}: \ \ && V={C_{n_e}\over (z-{\tilde z})^2},
\  n_e=1,2,3\,,  \nn\\ 
&& V \to 1\,, \ \ n_e=4\,, \nn\\ 
&& V\to 0\,, \ \     n_e=5,6,7\,,
\eea
where the coefficient is $C_{n_e}$ is independent of $N$ and depends
on $n_e$ in the following way:
\be
C_{n_e}=\fft{n_e(3n_e-8)}{4(n_e-4)^2}\,.\label{cne} 
\ee
The value of $C_{n_e}=\{-\textstyle{5\over {36}}, \ -\textstyle{1\over
{4}}, \ \textstyle{3\over 4}\}$ for $n_e=\{1,2,3\}$ respectively.

Thus barring the subtlety that the potential should not have any
additional local minimum, the spectra have the following universal
features.  For the cases $n_e=1,2,3$, the potentials give rise to
discrete spectra.  When $n_e=4$, the potential gives rise to a
continuous spectrum above a gap, whilst for $n_e=5,6,7$ the spectra
are continuous without any gap. (Note that had $C_{n_e}$ been less
than $ -\textstyle{1\over 4}$, the spectrum would have been unbounded
from below.  The case $n_e=2$ with $C_2=-\textstyle{1\over 4}$ is
precisely on the borderline.)

     Numerical plots of the various potentials are displayed in
Figures 1-7, for $n_e=1,\cdots ,7$, respectively.  In order to
facilitate the comparison of the potentials for a given $n_e$ but
different values of $N$, in Figures 1-3, the additive ambiguity in the
definition of $z$ has been fixed by requiring that ${\tilde z}_i=0$
for all $N$, and thus the ``horizon'' boundary corresponds to $z=0$
for all $N$. The $z$ coordinate was subsequently rescaled so that
$z=1$ corresponds to the boundary of AdS for all values of $N$.  In
Figures 4-7, the additive ambiguity in $z$ has been fixed such that
$z_N=0$, and thus $z=0$ corresponds to the AdS boundary for all $N$,
while $z\to -\infty$ is the horizon boundary.  The numerically
evaluated potentials, displayed in Figures 1-7, have no additional
local minima, thus confirming the nature of the spectra described
above.  For each value of $n_e$ the universal behavior at the horizon
boundary is apparent, and the potentials increase monotonically with
decreasing $N$.  Thus for each $n_e=1,2,3$ the discrete spectra have
an increasing gap between the discrete values of ${\cal Q}$ for the
M2-brane (solid lines), D3-brane (dotted lines), D4-D8-brane (dash
lines) and M5-brane (dash-dotted lines) respectively.  The rather
universal patterns for the spectra for each $n_e$ is also reflected
in the patterns for the charge distributions given in the
previous section. Note also that while the geometries with $n_e\le 4$
have ``naked'' singularities, nevertheless the spectra are regular
(and are {\it not unbounded} from below).
  
    A few additional comments are in order: The D3-brane case ($N=6$)
was analyzed in \cite{fgpw2}.\footnote{Note also in this case the
example with $n_e=4$ corresponds to the study of glue-ball ($0^{++}$)
spectra in Euclidean spinning-brane backgrounds with two angular
momenta, which were studied in \cite{rosf}. There, using the WKB
approximation, the first few discrete eigenvalues were found. However,
the present analysis shows that the spectrum becomes continuous for
higher excitations.} The integer $n$ used in \cite{fgpw2} is the same
as our $n$ for the Lorentz parameters, and thus our $n_e$ is related
to the $n$ in \cite{fgpw2} by $n_e=6-n$. We found no unbounded
spectrum for $n_e=1$.

\subsection{Spectrum with ``Lorentzian''  parameters}
\label{Lorentz}

Since the case with $n$ equal nonzero Lorentzian parameters
$\ell_1^2=\ell_2^2=\cdots =\ell_n^2>0$ is equivalent to the case with
$n_e=N-n$ Euclidean nonzero parameters, for the M2-brane ($N=8$),
D3-brane ($N=6$)
and M5-brane ($N=5$),\footnote{This is easily seen by performing a
coordinate transformation on either a solution (or the wave equation
(\ref{wave1})): $r^2\to r^2+\ell_i^2$. Thus both the distributions of
the brane configurations (see previous section) and the spectra in
these backgrounds are identical.} all these cases have already been
encompassed in the study above.

   On the other hand the D4-D8-brane does not have this symmetry (note
that in $D=6$, we have from (\ref{ndrel}) that $N=\textstyle{16\over
3}$, which is a non-integer) and the analysis for Lorentzian
parameters has to be done separately for these examples.  The wave
equation (\ref{wave2}) in this case can be cast in the form
(\ref{wavee}) where $y=r^2$ runs from 0 (at the horizon) to $\infty$
(the boundary of AdS) with the function $f(y)$ taking the form:
\begin{equation}
 f(y)=4 y^{\fft12(N-n)}\, (y+1)^{\fft12 n}\,,
\end{equation}
with $N=\ft{16}3$.  The quantity $f$ is again an increasing positive
function of $y$, with the asymptotic behavior
\bea
f(y) &\sim& y^{\fft12 N} \qquad \hbox{for $y \to \infty$}\ ,\nn\\
f(y) &\sim& y^{\fft12(N-n)} \qquad \hbox{for $y \to 0$}\,.
\eea
Here $n=1, \ldots, 4$ is the number of (equal) non-zero parameters
$\ell^2_i\ge 0$ whose values we took (without loss of generality) to be
$+1$, \ie  $\ell_1^2=\cdots \ell_n^2=1$.  (In the case of $\ell_i\ne
\ell_j$ the analysis of the spectra is qualitatively the
same as for the examples with ${\rm max}(\ell_i^2)=1$, and the other
parameters set to zero.)

Using the procedure, outlined above, to obtain the ``Schr\"odinger
potential'' $V(z)$, one arrives at the following features of the
potential $V(z)$ at the endpoints. The AdS boundary is again reached
as $y\to \infty$. At this boundary we have $z\to z'$, with
$z'$ finite, and the potential blows up quadratically as in (\ref{CN}),
with coefficient $C_N$ independent of $n$.  At the horizon boundary
($y\to 0$), we have $z\to {\tilde z}$ with ${\tilde z}$ finite for
$n=2,3,4$ and ${\tilde z}=-\infty$ for $n=1$.  At this boundary the
potential has the form
\bea
{\rm Horizon\  boundary}: \ \ && V\to 0 \  \,\ \     n=1\ \ ,  \nn\\ 
&&V={C_{\td n}\over (z-{\tilde z})^2},\  n=2,3,4\  .
\eea  
where $\td n=N-n$ and the coefficient is $C_{\td n}$ takes the same
form as (\ref{cne}) but with $n_e$ replaced by $\td n$.  Thus for the
D4-D8-brane with $N=\ft{16}3$, we find that when $n=1$ there is a
continuous spectrum without a gap, while the spectra for $n=2,3,4$ are
discrete, with the corresponding values $C_{\td
n}=[+\textstyle{15\over 4} , -\textstyle{21\over 100 }, \
-\textstyle{3\over 16} ]$ for the coefficients.  (Note again that
since $C_{\td n} >-\textstyle{1\over 4}$, the spectrum is always bounded
from below.)

The numerically evaluated potentials are displayed in Figures 1-4, for
$n=4,3,2,1$, respectively, as another (bold) solid line.  These 
potentials are quantitatively different in behavior from the respective
potentials with $n_e=1,2,3,4$.  Nevertheless, it appears that the cases with
$n$ and $n_e=5-n$ are qualitatively similar.

\section*{Note Added}

When this paper was nearing completion, the work \cite{BS} appeared, which
has some overlap with our results in the case of D3-branes.

\section*{Acknowledgements}

We thank D.~Freedman, S.~Giddings, K.~Pilch, and N.~Warner for useful
discussions.  The research of M.C.\ was in part by DOE grant
DOE-FG02-95ER40893, and also in part by the University of Pennsylvania
Research Foundation.  The research of S.S.G.\ was supported by the
Harvard Society of Fellows, and also in part by the NSF under grant
number PHY-98-02709, and by DOE grant DE-FGO2-91ER40654.  The research
of H.L.\ was supported in part by DOE grant DOE-FG02-95ER40893.  The
research of C.N.P.\ was supported by DOE grant DOE-FG03-95ER40917.
S.S.G. also thanks the Aspen Center for Physics for hospitality.  The
authors thank the ICTP and SISSA for hospitality during the initial
phases of the work.

 \begin{figure}
   \vskip0cm
\psfig{figure=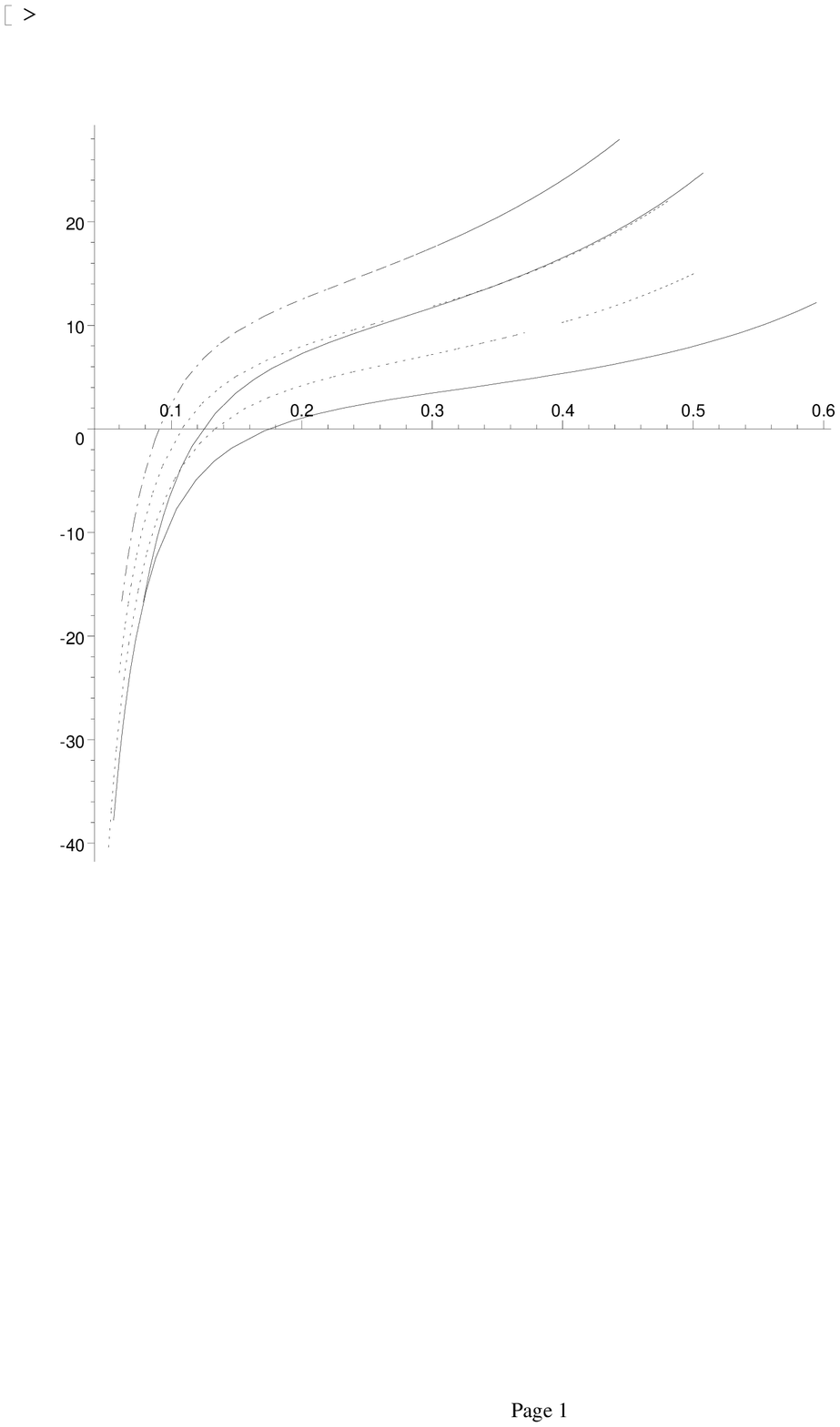,width=6in,angle=0,clip=}
   \vskip 2 cm
\caption{The Schr\" odinger  potentials  $V(z)$ with  $n_e=1$ Euclidean
 parameters are given for M2-branes, D3-branes, D4-D8-branes and
M5-branes, with successively increasing values
 of the potential.  The additional solid line represents 
the D4-D8-brane with $n=4$
 equal Lorentz parameters. The additive ambiguity for  
  $z$  is fixed in such a way that
 $z=0$ corresponds to the horizon boundary, and  $z$ is rescaled so
that $z=1$ is the boundary of AdS in all cases. }
\label{fig1}  \end{figure}
 
 \begin{figure}
   \vskip0cm
\psfig{figure=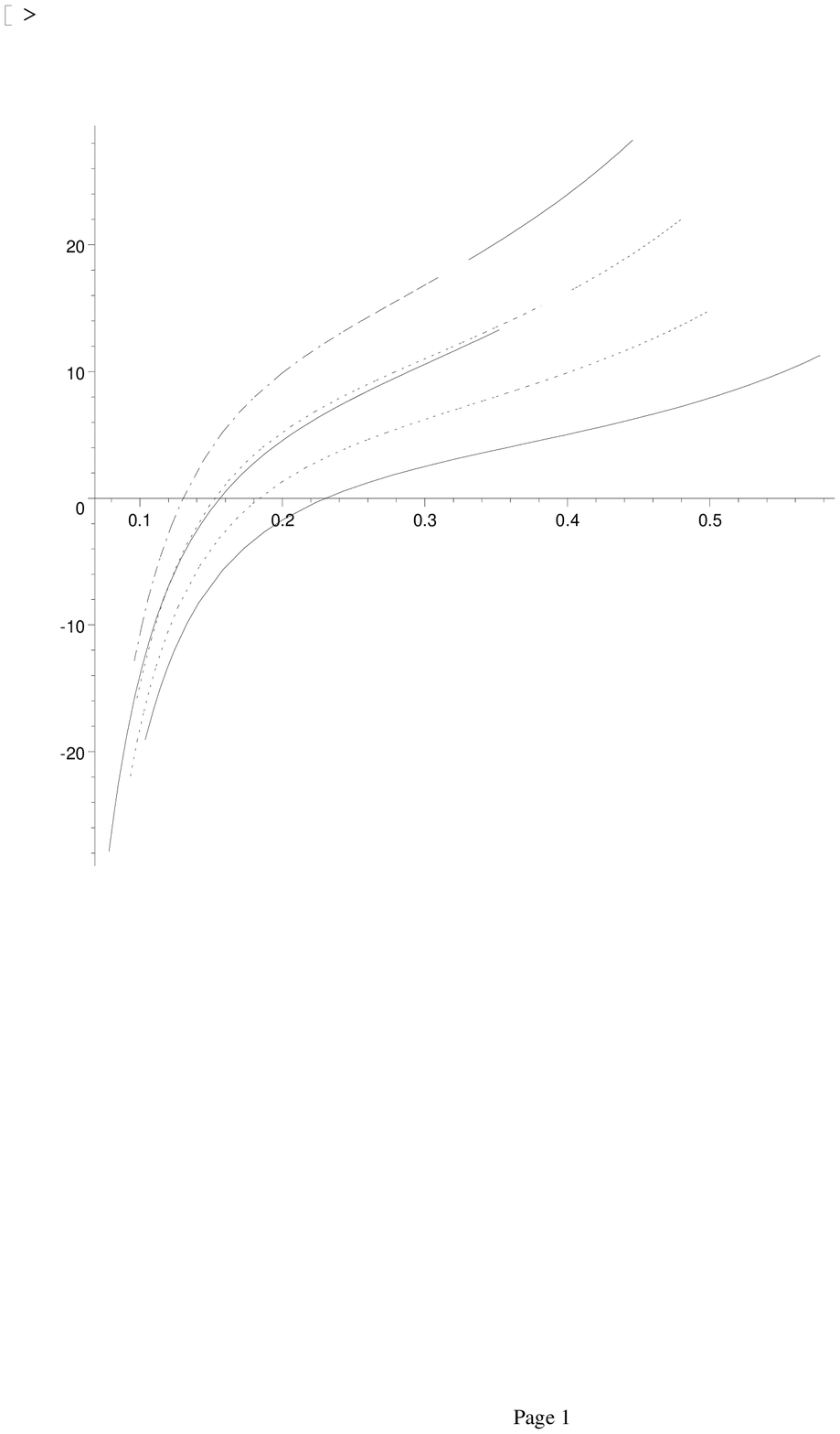,width=6in,angle=0,clip=}
   \vskip 2cm
\caption{The Schr\" odinger  potentials  $V(z)$ with  $n_e=2$ Euclidean
 parameters are given for M2-branes, D3-branes, D4-D8-branes and
M5-branes, with successively increasing values
 of the potential.   The additional  solid line represents the 
D4-D8-brane with $n=3$
 equal Lorentz parameters. The additive ambiguity for the 
  $z$  is fixed in such a way that
 $z=0$ corresponds to the horizon boundary, and  $z$ is rescaled so
 that $z=1$ is the boundary of AdS in all cases. }
\label{fig2}  \end{figure}

 \begin{figure}
   \vskip0cm
\psfig{figure=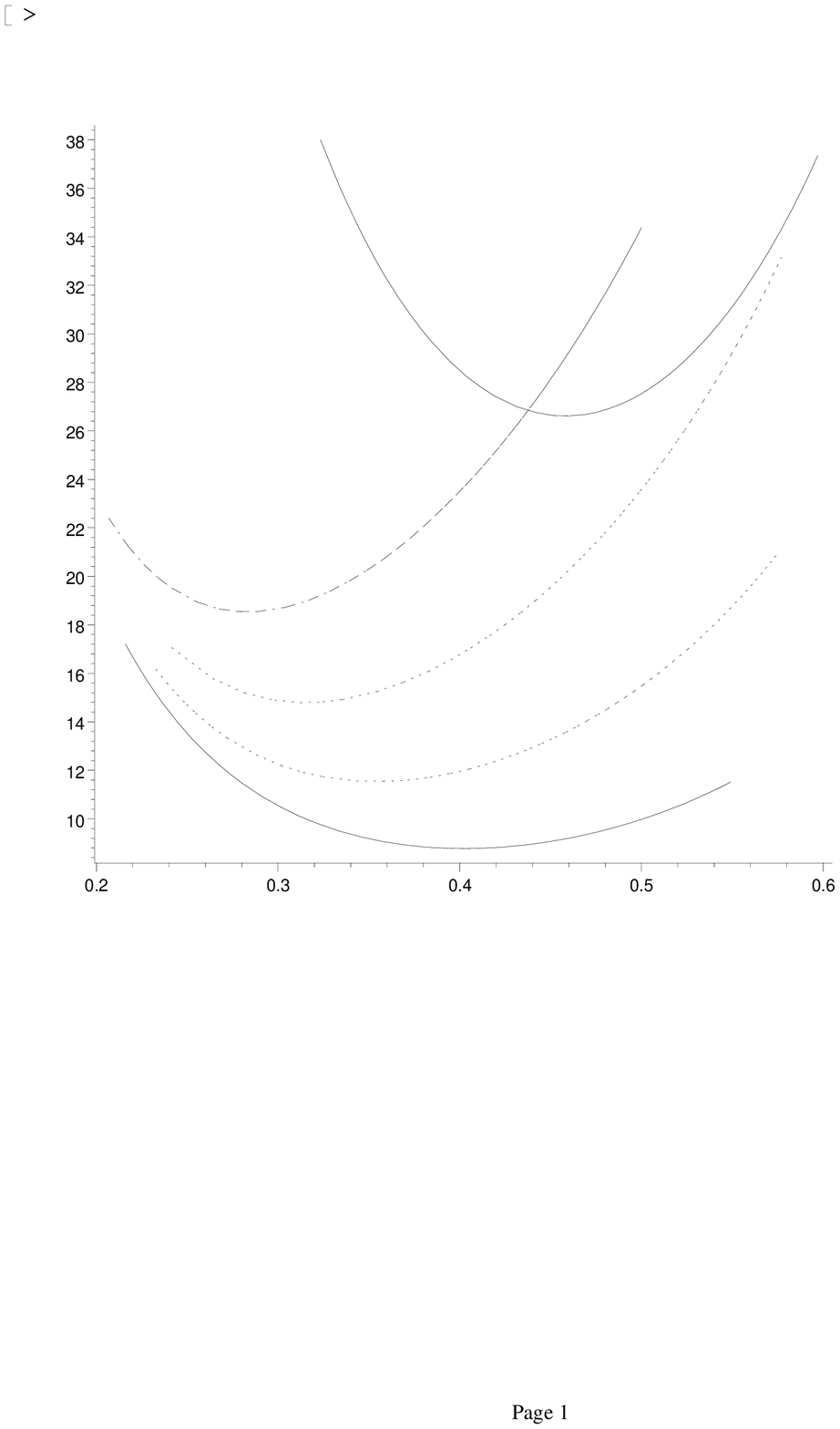,width=6in,angle=0,clip=}
   \vskip 2cm
\caption{The Schr\" odinger  potentials  $V(z)$ with  $n_e=3$ Euclidean
 parameters are given for M2-branes, D3-branes, D4-D8-branes and
M5-branes, with successively increasing values
 of the potential.  The additional solid line represents the 
D4-D8-brane with $n=2$ equal Lorentz parameters. The additive ambiguity for  
  $z$  is fixed in such a way that
 $z=0$ corresponds to the horizon boundary, and  $z$ is rescaled so
 that  $z=1$ is the boundary of AdS in all cases. }
\label{fig3}  
\end{figure}

\begin{figure}
   \vskip0cm
\psfig{figure=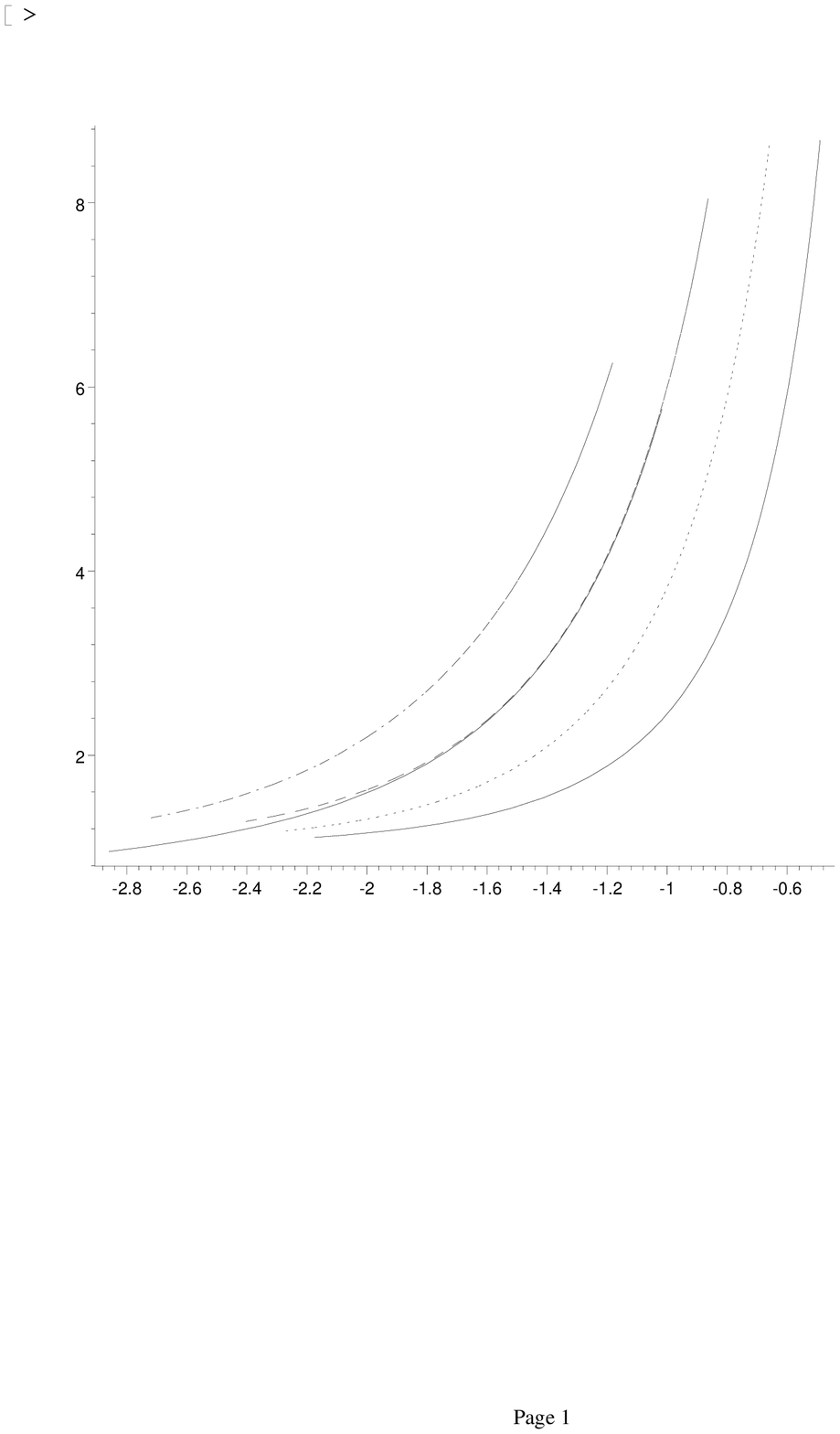,width=6in,angle=0,clip=}
   \vskip 2cm
\caption{The Schr\" odinger  potentials  $V(z)$ with  $n_e=4$ Euclidean
 parameters are given for M2-branes, D3-branes, D4-D8-branes and
M5-branes,  with successively increasing values
 of the potential.  The additional solid line represents the 
D4-D8-brane with $n=1$ Lorentz parameter.
 $z=0$ corresponds to the  AdS boundary. $V\to 1$ for the $n_e=4$
cases and $V\to 0$ for $n=1$.}
\label{fig4} 
 
 \end{figure}
\begin{figure}
   \vskip0cm
\psfig{figure=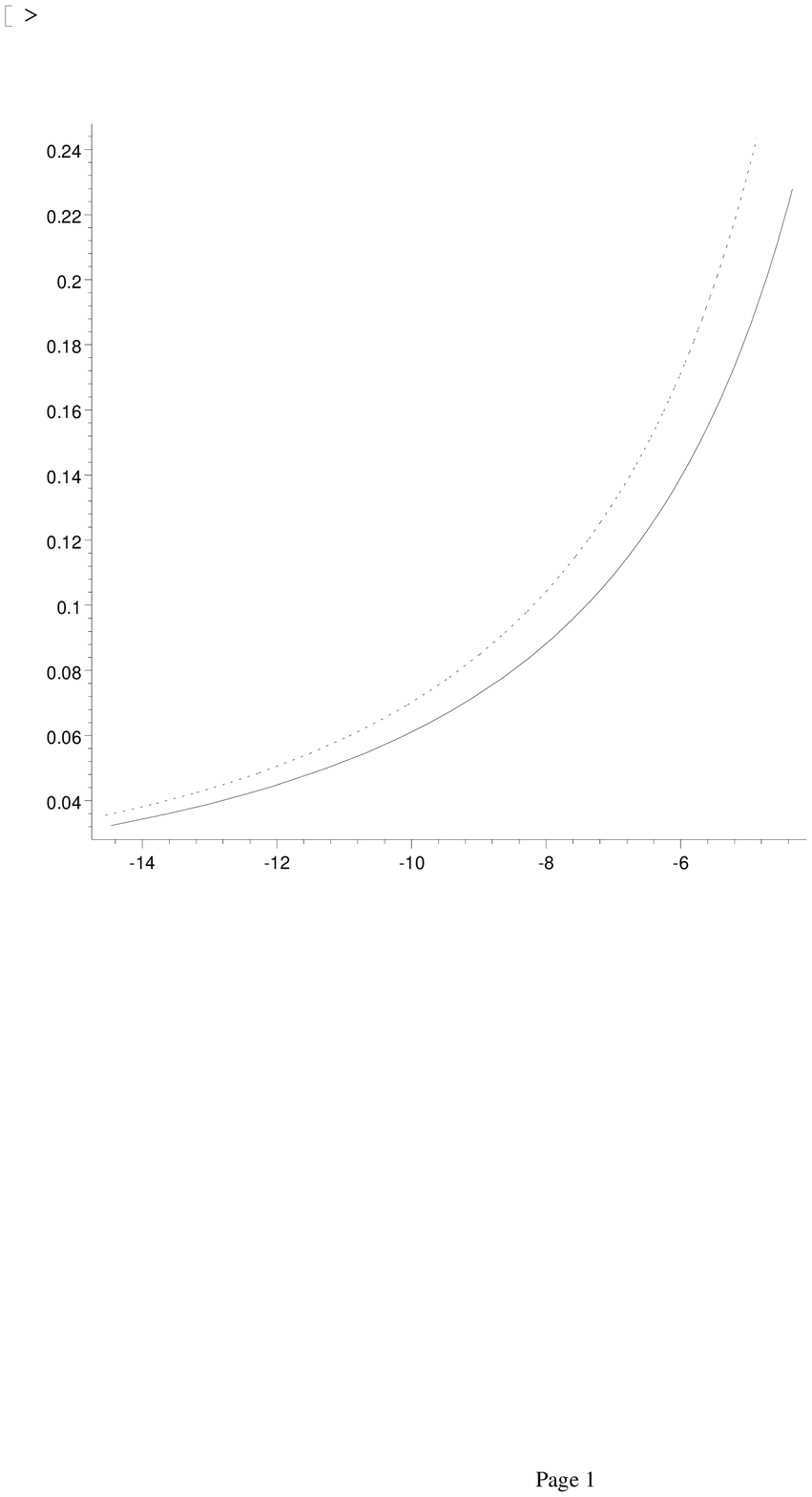,width=6in,angle=0,clip=}
   \vskip 2cm
\caption{The Schr\" odinger  potentials  $V(z)$ with  $n_e=5$ Euclidean
 parameters are given for M2-branes and D3-branes, with successively 
increasing values
 of the potential. The AdS boundary is at $z=0$. }
\label{fig5}  \end{figure}
\begin{figure}
   \vskip0cm
\psfig{figure=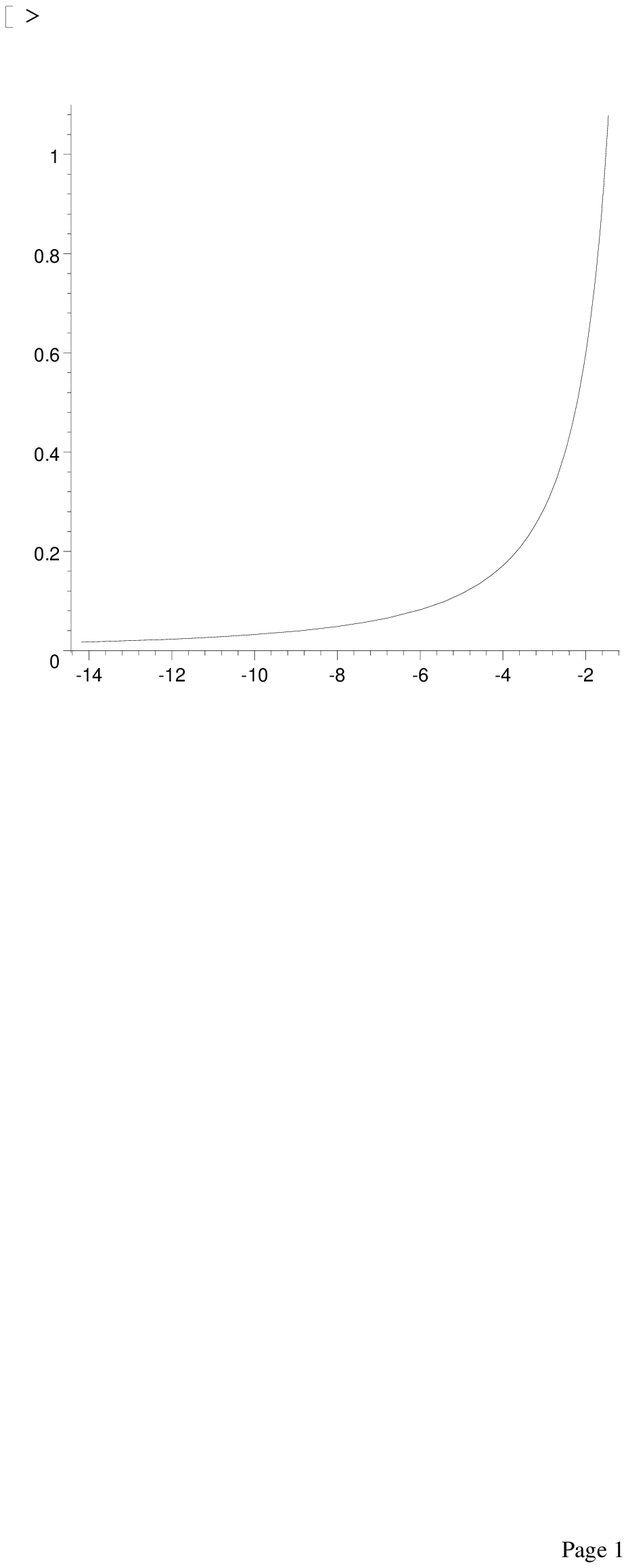,width=6in,angle=0,clip=}
   \vskip 2cm
\caption{The Schr\" odinger  potential  $V(z)$ with  $n_e=6$ Euclidean
 parameters is given for the M2-brane.  The AdS boundary is at $z=0$.  }
\label{fig6}  \end{figure}\begin{figure}
   \vskip0cm
\psfig{figure=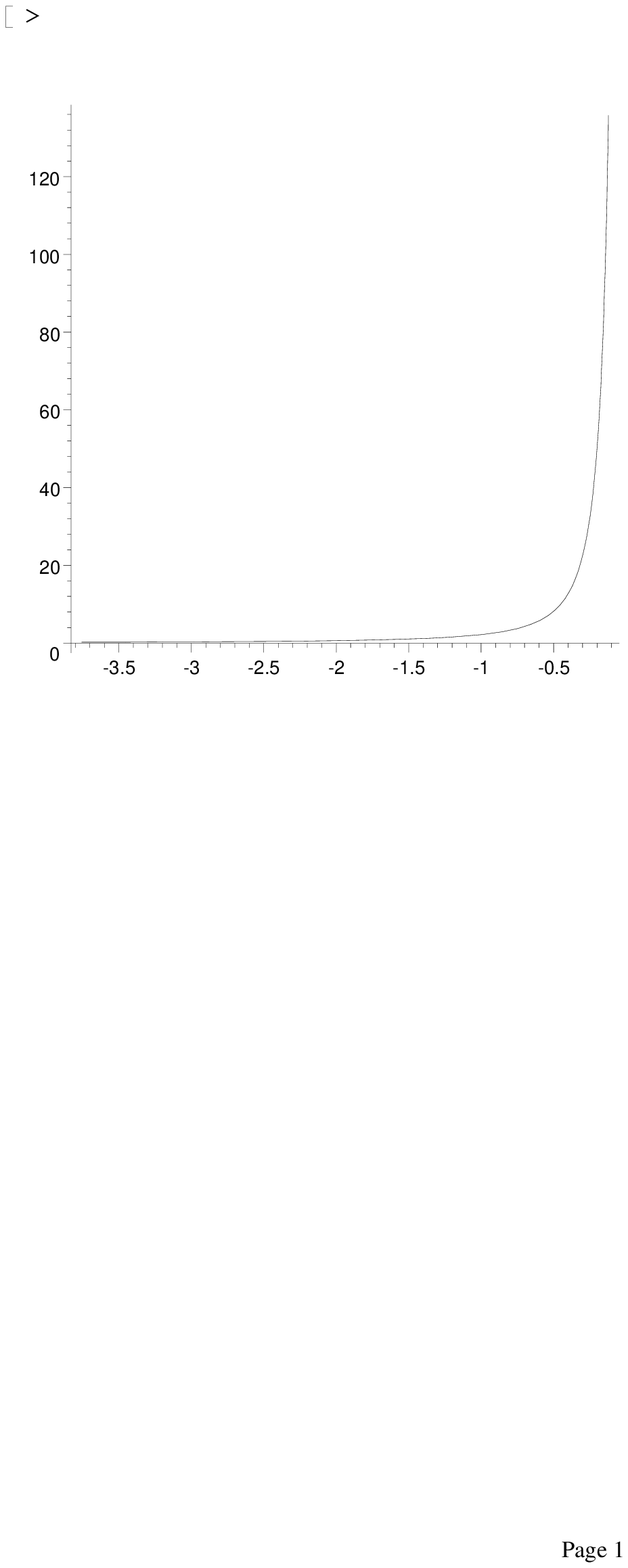,width=6in,angle=0,clip=}
   \vskip 2 cm
\caption{The Schr\" odinger  potential $V(z)$ with  $n_e=7$ Euclidean
 parameters is given for the M2-brane.  The AdS  boundary is at $z=0$. }
\label{fig7}  \end{figure}

\end{document}